\documentclass[twocolumn, tighten, twocolappendix]{aastex631_mod}
\usepackage[T1]{fontenc}
\usepackage{microtype}
\usepackage{newtxtext}
\usepackage[varvw]{newtxmath}
\usepackage{mathtools}

\hypersetup{pdfauthor={Errani et al.},
            pdftitle={},
            pdfkeywords={},
            bookmarksnumbered=true}

\newcommand{\rmx}{r_\mathrm{mx}}

\newcommand{\Vmx}{V_\mathrm{mx}}

\newcommand{\kms}{\mathrm{km\,s^{-1}}}
\newcommand{\ms}{\mathrm{m\,s^{-1}}}
\newcommand{\Rh}{R_\mathrm{h}}

\newcommand{\rh}{r_\mathrm{h}}

\newcommand{\diff}{\mathrm{d}}
\newcommand{\Msol}{\mathrm{M_{\odot}}}

\newcommand{\Vc}{V_\mathrm{c}}

\newcommand{\kpc}{\mathrm{kpc}}
\newcommand{\pc}{\mathrm{pc}}
\newcommand{\Gyr}{\mathrm{Gyr}}
\newcommand{\Myr}{\mathrm{Myr}}
\newcommand{\rperi}{r_\mathrm{peri}}

\newcommand{\Torb}{T_\mathrm{orb}}
\newcommand{\rapo}{r_\mathrm{apo}}
\newcommand{\mxzero}{_\mathrm{mx0}}
\newcommand{\dex}{\mathrm{dex}}

\newcommand{\sigmalos}{\sigma_\mathrm{los}}

\newcommand{\g}{\mathrm{g}}
\newcommand{\E}{\mathcal{E}}
\newcommand{\plus}[1] {^{\mathmakebox[\widthof{$^-$}][c]{+}#1}}
\newcommand{\minus}[1]{_{\mathmakebox[\widthof{$^-$}][c]{-}#1}}

\graphicspath{{./}{./Figures/}}

\begin{document}
\shorttitle{UMa3/U1: The Darkest Galaxy Ever Discovered?}
\shortauthors{Errani et al.}
\title{Ursa Major III/UNIONS 1: The Darkest Galaxy Ever Discovered?}

\author{Rapha\"el Errani}
\affiliation{McWilliams Center for Cosmology, Department of Physics, Carnegie Mellon University, Pittsburgh, PA 15213, USA}
\affiliation{Universit\'e de Strasbourg, CNRS, Observatoire Astronomique de Strasbourg, UMR 7550, F-67000 Strasbourg, France}
\email{errani@cmu.edu}

\author{Julio F. Navarro}
\affiliation{Department of Physics and Astronomy, University of Victoria, Victoria, BC, V8P 5C2, Canada}

\author{Simon E. T. Smith}
\affiliation{Department of Physics and Astronomy, University of Victoria, Victoria, BC, V8P 5C2, Canada}

\author{Alan W. McConnachie}
\affiliation{NRC Herzberg Astronomy and Astrophysics, 5071 West Saanich Road, Victoria, BC, V9E 2E7, Canada}
\affiliation{Department of Physics and Astronomy, University of Victoria, Victoria, BC, V8P 5C2, Canada}

\received{2023 October 30}
\revised{2024 January 7}
\accepted{2024 January 23}

\begin{abstract}
The recently discovered stellar system Ursa Major III/UNIONS 1 (UMa3/U1) is the faintest known Milky Way satellite to date. With a stellar mass of $16\plus{6}\minus{5}\,\Msol$ and a half-light radius of $3\pm1\,\pc$, it is either the darkest galaxy ever discovered or the faintest self-gravitating star cluster known to orbit the Galaxy. Its line-of-sight velocity dispersion suggests the presence of dark matter, although current measurements are inconclusive because of the unknown contribution to the dispersion of potential binary stars. We use $N$-body simulations to show that, if self-gravitating, the system could not survive in the Milky Way tidal field for much longer than a single orbit (roughly $0.4\,\Gyr$), which strongly suggests that the system is stabilized by the presence of large amounts of dark matter. If UMa3/U1 formed at the center of a ${\sim}10^{9}\, \Msol$ cuspy LCDM halo, its velocity dispersion would be predicted to be of order ${\sim}1\,\kms$. This is roughly consistent with the current estimate, which, neglecting binaries, places $\sigma_{\rm los}$ in the range $1$ -- $4\,\kms$. Because of its dense cusp, such a halo should be able to survive the Milky Way tidal field, keeping UMa3/U1 relatively unscathed until the present time. This implies that UMa3/U1 is plausibly the faintest and densest dwarf galaxy satellite of the Milky Way, with important implications for alternative dark matter models and for the minimum halo mass threshold for luminous galaxy formation in the LCDM cosmology. Our results call for multi-epoch high-resolution spectroscopic follow-up to confirm the dark matter content of this extraordinary system.
\end{abstract}

\keywords{Cold dark matter (265); Dwarf spheroidal galaxies (420); Low surface brightness galaxies (940); the Milky Way (1054); N-body simulations (1083); Star clusters (1567); Tidal disruption (1696)}

\defcitealias{EN21}{EN21}
\defcitealias{EP20}{EP20}
\defcitealias{Smith2024}{S+24}

\section{Introduction}
\label{SecIntro}
Ultrafaint dwarf galaxies (UFDs) are stellar systems with $M_\star<10^5\,\Msol$, fainter than many globular clusters (GCs) but gravitationally bound by the presence of large amounts of dark matter. They constitute direct probes not only of the formation mechanisms that govern the extreme faint-end of the galaxy luminosity function, but also of the structure of low-mass dark matter halos and, indirectly, of the nature of dark matter \citep[see, e.g.,][for recent reviews]{Bullock2017_Review,Simon2019Review,Sales2022}.

The overall abundance of UFDs reflects the number of low-mass dark matter halos able to harbor luminous galaxies, placing important constraints on models where the physical nature of dark matter leads to the suppression of low-mass halos, such as in ``warm dark matter'' \citep[WDM; e.g.][]{Bode2001, Lovell2014} or ``fuzzy dark matter'' \citep[FDM; e.g.][]{Hu2000} models. For cold dark matter (CDM) models, where the number of low-mass halos is expected to be overwhelmingly larger than the number of UFDs, the abundance of faint systems probes the mass threshold between halos that remain ``dark'' (starless) and those massive enough for luminous galaxy formation to proceed \citep{Simon2007,Ferrero2012,Penarrubia2012,Fattahi2018}.

This threshold is still being actively discussed, with some studies suggesting a relatively high virial\footnote{Virial quantities are identified by a ``200'' subscript and defined at or within the virial radius, $r_{200}$, which encloses a mass a mean average density equal to $200$ times the critical density for closure. At $z=0$, $\rho_\mathrm{crit} =  3 H_0^2 / (8 \pi G)$, with $H_0 = 67\kms \mathrm{Mpc}^{-1}$ \citep{Planck2020}. \label{footnote:virial}} halo mass threshold \citep[${\sim} 10^9\, \Msol$;][]{Benitez-Llambay2020, Pereira-Wilson23}, determined primarily by the ability of hydrogen to cool in  halos photoheated by the ambient UV background \citep{Efstathiou1992,Quinn1996,Gnedin2000}, and other studies arguing for a much lower mass threshold, in order to accommodate the sheer number of observed UFDs plus those still likely missing from our currently incomplete inventory of Milky Way satellites \citep[e.g.,][and references therein]{Nadler2021}.

In addition, because UFDs are physically small and heavily dark-matter-dominated, they probe the innermost regions of their dark matter halos, where competing dark matter models make differing predictions. CDM models predict ``cuspy'' density profiles \citep{Navarro1996a,Navarro1997}, which imply high average dark matter densities for UFDs. Cuspy halos are remarkably resilient to tidal stripping \citep[][hereafter \citetalias{EN21}]{Penarrubia2008, Penarrubia2010, Amorisco2021, EN21}, and may host ``micro galaxies'' even after undergoing substantial tidal mass loss \citep[][hereafter \citetalias{EP20}]{EP20}.

On the other hand, self-interacting dark matter (SIDM) or FDM models generally predict much lower dark matter densities, as a result of the inward energy transfer driven by self-interactions \citep[SIDM --][]{Colin2002,Zavala2013, Tulin2018} or of the quantum pressure support arising from the uncertainty principle \citep[FDM --][]{Hu2000, Goodman2000, Burkert2020,Ferreira2021}. 

Available data indicate that UFDs are, indeed, quite dense \citep[see, e.g.,][]{Simon2019Review, BattagliaNipoti2022Review}. This result has led recent SIDM modeling to consider much higher interaction cross sections than envisioned in earlier work \citep{Silverman2023}, in an attempt to reconcile observations with SIDM halos whose inner densities have been gravothermally enhanced by ``core collapse'' \citep{Balberg2002, Nishikawa2020, Turner2021, Correa2022, Zeng2022, Zeng2023}. The same result places strong constraints on FDM models too, and suggests that earlier lower-mass bounds for ultra-light particles based on the ``classical'' dwarf spheroidal (dSphs) satellites of the Milky Way should be drastically revised \citep{Safarzadeh2020}.

Finally, UFDs are ideal laboratories for studying in unprecented detail the heavy element enrichment process driven by recurring episodes of star formation. Some of these galaxies are so faint and so metal poor that the abundance pattern of individual stars may well reveal the nucleosynthetic yields of individual Population III supernovae or other explosive stellar events in these chemically pristine systems \citep{Frebel2015, Hansen2017, Ji2019_Gru1, Marshall2019}.

These traits make UFDs highly valuable, and have given impetus to a number of specialized UFD searches using resolved stars in widefield photometric surveys. Because they are so faint, UFDs are elusive objects that barely stand out against the foreground of Galactic stars and the background of distant galaxies.

UFD candidates are typically identified by matched-filter techniques, which pinpoint clumps of old stars at a common distance \citep{Koposov2008,Walsh2009,Martin2013,Drlica-Wagner2015}.
These clumps are then followed up with deeper photometry and spectroscopy to enable a full characterization of the system. When available, proper motions from the Gaia mission \citep{Simon2018, Pace2019, McConnachieVenn2020b, McConnachieVenn2020a,  LiHammer2021} can help to aid the discovery process, but at the expense of being applicable only to relatively nearby systems, given Gaia's relatively shallow depth compared to contemporary widefield, digital photometric surveys.

Distinguishing dark-matter-dominated UFDs from self-gravitating faint star clusters is the final, perhaps most difficult hurdle, one that can only be fully overcome by securing multi-epoch line-of-sight velocities to test whether the system is a self-gravitating star cluster or a UFD bound by the presence of dark matter \citep{WillmanStrader2012}.

For the faintest and most distant candidates, this is a most challenging task, given the few stars bright enough to obtain spectra for, the limited precision of the individual radial velocities, the uncertainties from low-number statistics \citep{Laporte2019}, and the possibility that binary stars may lead to inflated values of the velocity dispersion, confusing the interpretation \citep{McConnachieCote2010, MinorQuinn2010}.

Although a few dozen candidate\footnote{We colloquially refer to these systems as ``glorfs'' -- i.e., either a dwarf galaxy or a GC.} UFDs are currently known, with total luminosities in the range $+1>M_\mathrm{V}>-3$ and sizes in the range $1$ -- $20\,\pc$ \citep[e.g.][]{Munoz2012, Balbinot2013, Torrealba2019clusters, Mau2020,  Cerny2023DECam, Cerny2023Delve}, very few of those have been conclusively identified as dark-matter-dominated UFDs because of the difficulties listed above.

The most singular of these candidates is Ursa Major III/UNIONS 1 (hereafter UMa3/U1), an astonishingly faint ($M_\mathrm{V}\sim +2.2$) stellar system with a projected half-light radius of only $\Rh=3\pm 1$ pc recently discovered by \citet{Smith2024}, hereafter \citetalias{Smith2024}. UMa3/U1 orbits the Milky Way on an inclined, ``halo-like'' orbit with pericenter $\rperi\approx 13\,\kpc$ and apocenter $\rapo \approx 30\,\kpc$. Although comparable in size to GCs, it is ${\sim}10$ times fainter than the faintest known GC to date \citep{Inman1987, Koposov2007}.  It is also at least ${\sim}10$ times fainter and ${\sim}5$ times smaller than the faintest confirmed UFDs \citep{Belokurov2009,Geha2009,Martin2016_Dra2}.

\citetalias{Smith2024} report a velocity dispersion for UMa3/U1 of $\sigmalos=3.7\plus{1.4}\minus{1.0}\,\kms$ on the basis of 11 likely member stars, which would imply a mass-to-light ratio of several thousands. The authors do, however, caution that the estimate of $\sigmalos$ is very sensitive to the inclusion (or exclusion) of specific member stars: removing a single star (the largest outlier in velocity, perhaps a binary star) drops the estimate to $\sigmalos=1.9\plus{1.4}\minus{1.1}\,\kms$. Removing a second outlier from the sample results in a formally unresolved velocity dispersion, consistent with the extremely small value expected if UMa3/U1 were a self-gravitating star cluster ($\sigmalos \sim 50$ m/s).
Should $\sigmalos$ prove much higher than this value, it would imply that UMa3/U1 is the faintest, or ``darkest'', galaxy ever discovered.

Because of the sensitivity of the  $\sigmalos$ estimate to the two outliers, and because of the lack of repeat velocity measurements (needed to rule out the undue influence of binary stars), \citetalias{Smith2024} are unable to ascertain the true nature of  UMa3/U1. It is clear, however, that regardless of its nature, this system is truly exceptional.

We present here a simple argument in favor of the interpretation of UMa3/U1 as a genuine dark-matter-dominated UFD. The argument relies on the fact that should UMa3/U1 be a self-gravitating star cluster lacking dark matter, its average density would be comparable to the mean density of the Galaxy at the pericenter of its orbit. As such, it could not survive long on its current orbit, which, given the short orbital time of UMa3/U1 around the Galaxy, seems extremely unlikely.

We elaborate on this idea in Sec.~\ref{SecSG}, where we use $N$-body simulations to model the tidal evolution of UMa3/U1, under the assumption that it is a self-gravitating star cluster.
In Sec.~\ref{SecDM}, we model UMa3/U1 as a dark-matter-dominated \emph{micro galaxy} and show that its estimated velocity dispersion is consistent with that expected if UMa3/U1 inhabits a cuspy ${\sim} 10^{9}\, \Msol$ CDM halo.
Finally, we summarize and discuss our main conclusions in Sec.~\ref{SecConc}.

\section{Self-Gravitating (SG) Model }
\label{SecSG}
Assuming that UMa3/U1 is a self-gravitating star cluster, we adopt a simple model where its density profile is approximated by a spherical exponential distribution,
\begin{equation}
 \rho_\star(r) = \rho_0 ~ \exp(-r/r_\star)~,
 \label{eq:3DExp}
\end{equation}
with a scale radius $r_\star$ and a total stellar mass $ M_\star = 8 \pi \, \rho_0 \,  r_\star^3$. The 3D and 2D half-mass radii are related to the scale radius through $\rh \approx 2.67~ r_\star$ and $\Rh \approx 2.02~ r_\star$, respectively. 
Under the assumption of dynamical equilibrium, the line-of-sight velocity dispersion may be computed from the projected virial theorem \citep{Amorisco2012,EPW18},
\begin{equation}
  \langle \sigmalos^2 \rangle = \frac{5}{96} \frac{G M_\star}{r_\star}~.
\end{equation}
Using the parameters estimated by \citetalias{Smith2024}, $M_\star = 16\plus{6}\minus{5}\,\Msol$ and $\Rh = (3\pm1)\,\pc$, we estimate, for the self-gravitating case, a line-of-sight velocity dispersion of
\begin{equation}
\label{eq:selfgrav_disp}
 \sigmalos \equiv \langle \sigmalos^2 \rangle^{1/2} = 49\plus{14}\minus{11} \, \ms~. 
\end{equation}
The uncertainties above are estimated from a Monte Carlo sample, taking into account the asymmetric measurement uncertainties on $M_\star$ and $\Rh$.
Note that a velocity dispersion this small is well below what \citetalias{Smith2024} could measure, given their observational setup.

Confirming that the (virial) velocity dispersion of UMa3/U1 is indeed of the order of a few $\kms$, as suggested from the dispersion estimate of  \citetalias{Smith2024} from the 10- and 11-member samples (see Sec.~\ref{SecIntro}), would rule out conclusively and convincingly the possibility that this system is a self-gravitating star cluster.

\subsection{Tidal Evolution}
\subsubsection{N-body Models}
We use $N$-body simulations to analyze the evolution of the self-gravitating UMa3/U1 model described above in the Milky Way gravitational potential.
With a total stellar mass of $M_\star =16\plus{6}\minus{5}\,\Msol$, UMa3/U1 likely contains only a few dozen stars (\citetalias{Smith2024} estimate $N_\star \sim  21\plus{6}\minus{5}$ stars brighter than $23.5\,\mathrm{mag}$). The system is therefore intrinsically \textit{collisional}, giving rise to a complex internal dynamical evolution that depends on the initial stellar mass function, the fraction of binary stars, the number of potential stellar remnants like neutron stars and black holes, and the exact stochastic realization of the underlying distribution function. None of these initial properties are well understood for faint stellar systems. A full exploration of this parameter space is hence, at the present day, impractical at best. With these caveats in mind, we model in this section UMa3/U1 as a \textit{collisionless} system, which should be enough to broadly illustrate the dynamical evolution of UMa3/U1 in the Milky Way potential.

The progenitor of UMa3/U1 is modeled as an $N$-body realization of an exponential sphere with $10^6$ particles with isotropic velocity dispersion, generated using the code described in \citet{EP20}, available online\footnote{\label{footnote:nbopy}\url{https://github.com/rerrani/nbopy}}.
We explore models with initial masses in the range $16 \leq M_\star / \Msol \leq 160$. All models have the same initial (2D) half-light radius of $\Rh = 3\,\pc$.

For the Milky Way halo, we assume the static, analytical potential from \citet{EP20}. The potential includes a thick and a thin axisymmetric \citet{miyamoto1975} disk ($M_\mathrm{thick}=2.0\times10^{10}\,\Msol$, $M_\mathrm{thin}=5.9\times10^{10}\,\Msol$, radial scale lengths $a_\mathrm{thick} = 4.4\,\kpc$, $a_\mathrm{thin}=3.9\,\kpc$, and vertical scale lengths $b_\mathrm{thick} = 0.92\,\kpc$, $b_\mathrm{thin}=0.31\,\kpc$ for the thick and thin disc, respectively), a spherical \citet{hernquist1990} bulge ($M_\mathrm{b}=2.1\times10^{10}\,\Msol$ and scale length $a_\mathrm{b}=1.3\,\kpc$), and a Navarro-Frenk-White \citep[NFW;][]{Navarro1996a,Navarro1997} dark matter halo ($M_\mathrm{200} = 1.15 \times 10^{12}\,\Msol$, $r_\mathrm{200}=192\,\kpc$, and concentration $c=9.5$), with parameters chosen to approximate the circular velocity curve of \citet{McMillan2011}. In practice, the adopted Galactic potential is not too different, in the regions of interest, from that of a singular isothermal sphere with constant circular velocity, $\Vc\approx240\,\kms$. See Appendix~\ref{Appendix_orbits} for a wider exploration of potentials and orbits.

The integration is performed using the particle mesh code \textsc{superbox} \citep{Fellhauer2000}. This code employs two cubic grids of $128^3$ cells comoving with the $N$-body model with resolutions of $0.04\,\pc$ and $0.4\,\pc$, respectively, as well as a fixed $128^3$ cell grid containing the entire simulation volume, with a lower resolution of $\approx1.6\,\kpc$.

\subsubsection{Tidal Disruption Timescales}
\begin{figure}
\begin{center}
 \includegraphics[width=\columnwidth]{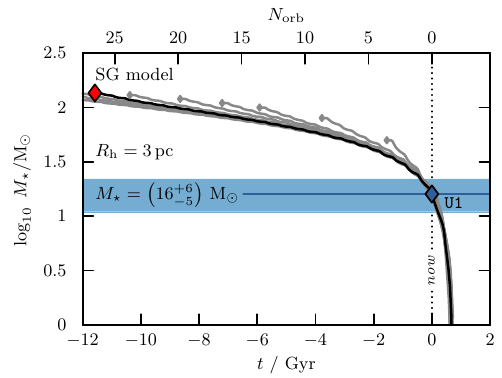} 
\end{center}
 \caption{Mass evolution of self-gravitating (SG) $N$-body models for the UMa3/U1 stellar system. All models have an initial (2D) half-light radius of $\Rh = 3\,\pc$. A blue band shows the 1-$\sigma$ measurement uncertainty around the current mass of UMa3/U1. The evolution of an example model with initial mass $M_\star = 136\,\Msol$ (the red diamond symbol) is highlighted in black. Note that all models, independently of their initial mass, fully disrupt within ${\sim}0.6\,\Gyr$ from now, suggesting that if UMa3/U1  is a self-gravitating object, then we observe it at a very special point in time in its evolution.}
 \label{fig:SG_mass_evolution}
\end{figure}

We begin by simply evolving the SG UMa3/U1 model forward from its observed present-day position and proper motions, as listed by \citetalias{Smith2024}. Using the \citet{SB2010} solar velocity with respect to the local standard of rest results in an orbit with a pericenter of $\rperi \approx 13\,\kpc$, and apocenter of $\rapo \approx30\,\kpc$. The total bound mass is shown, as a function of time, in Figure~\ref{fig:SG_mass_evolution}, where the blue diamond symbol (along the vertical dotted line labeled ``now'') represents the present-day configuration of UMa3/U1. As is clear from this figure, the system fully disrupts in less than two radial orbital times (with $\Torb=0.4\,\Gyr$ in the \citetalias{EP20} potential). This result holds for all orbits compatible with the observed position and velocity of UMa3/U1; see Appendix~\ref{Appendix_orbits}.

The short survival time of the SG UMa3/U1 model is not surprising, for its mean density is comparable to the mean density of the Milky Way inside the pericenter of the orbit:
\begin{equation}
 \bar \rho_\mathrm{peri} = 3/(4\pi G) ~ (V_\mathrm{peri}^2/\rperi^2) \approx 1.9 \times 10^7\, \Msol\, \kpc^{-3} ~,
\end{equation}
where $V_\mathrm{peri}\approx240\,\kms$ \citep{Huang2016, Eilers2019} is the circular velocity of the Milky Way at $\rperi$. For comparison, the mean density of the SG UMa3/U1 model within its 3D half-light radius is
\begin{equation}
\bar \rho_{\star\mathrm{h}} = 3\,M_\star/(8\pi \rh^3) = 2.9\plus{7.2}\minus{1.8} \times 10^7\, \Msol\, \kpc^{-3}~,
\end{equation}
where the quoted uncertainty takes into account the asymmetric measurement uncertainties on stellar mass $M_\star$ and half-light radius.

\begin{figure}
\begin{center}
 \includegraphics[width=\columnwidth]{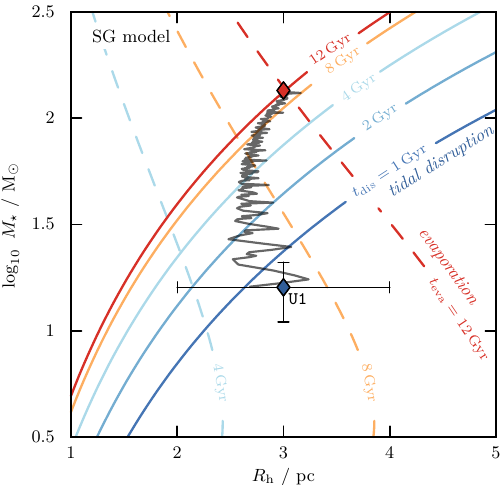} 
\end{center}
 \caption{The same as Fig.~\ref{fig:SG_mass_evolution}, but for the tidal evolution in the mass--size plane. Numerical estimates of the tidal disruption ($t_\mathrm{dis}$) and evaporation ($t_\mathrm{eva}$, Eq.~\ref{eq:evaporation_time}) timescales are shown by the solid and dashed colored curves, respectively. The present-day mass and size of UMa3/U1 are shown as the blue diamond with $1\,\sigma$ error bars. A gray curve shows the evolution of the example model highlighted in Fig.~\ref{fig:SG_mass_evolution}.  }
 \label{fig:SG_IC_systematics}
\end{figure}

To complete the analysis, we modify the initial mass of the SG UMa3/U1 model to ensure that it survives for a longer period of time on the same orbit, and adjust it to match, at present, the observed properties of UMa3/U1. The initial conditions are generated by first integrating the orbit backward in time. For the model to match today's properties after evolving for, say, $12$ Gyr (i.e., ${\sim} 27$ full orbits), it must have been substantially more massive/denser, as indicated by the red diamond symbol in Fig.~\ref{fig:SG_mass_evolution}. Tidal mass loss little affects the size of the bound remnant, so a model with a similar half-mass radius but an initial mass of ${\sim} 136\, \Msol$ could, in principle, have been the progenitor of today's UMa3/U1. The evolution of this progenitor in the mass-radius plane is shown in Fig.~\ref{fig:SG_IC_systematics}.

The colored curves in Fig.~\ref{fig:SG_IC_systematics} are fitted to disruption times measured in the simulations. As shown in \citet{ENPFI2023}, on a given orbit, the disruption times depend mainly on the initial density contrast between progenitor and host, measured at pericenter.
Intermediate values of the initial mass (i.e., in the range $16 \leq M_\star / \Msol \leq 160$) lead, as expected, to intermediate survival times, as shown by the gray lines/diamonds in Fig.~\ref{fig:SG_mass_evolution}. Still, all of these models disrupt fully in less than ${\sim}0.6\,\Gyr$ from now. Given that stars in UMa3/U1 are likely $\gtrsim 11$ Gyr old \citepalias{Smith2024}, it would require quite the coincidence to discover UMa3/U1 just as we witness its final orbit around the Milky Way.

\subsubsection{Tidal Debris}
\begin{figure}
\begin{center}
 \includegraphics[width=\columnwidth]{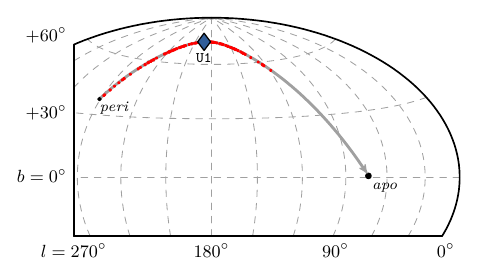} 
 
 \includegraphics[width=\columnwidth]{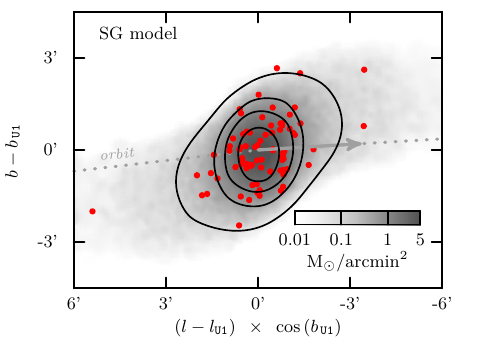}
\end{center}
 \caption{Top panel: orbit of UMa3/U1 in Galactic coordinates. The red points show a Monte Carlo sample of stripped stars drawn from the $N$-body model highlighted in Fig.~\ref{fig:SG_mass_evolution}. Bottom panel: surface density map of the self-gravitating (SG) model (grayscale). The orbit is shown using a dotted curve, with an arrow indicating the direction of motion along the orbit. }
 \label{fig:selfgrav_lb}
\end{figure}

However unlikely the SG UMa3/U1 model might be, a robust prediction that may be scrutinized observationally is the presence of tidal debris along the orbit. Given the extremely low velocity dispersion of the progenitor, the debris should align along a thin stream, as depicted by the red dots in the top panel of Fig.~\ref{fig:selfgrav_lb}. The particular realization shown in this figure corresponds to the $12\,\Gyr$ old progenitor identified by the red diamond in Fig.~\ref{fig:SG_mass_evolution}, but the configuration would be similar for other massive progenitors. 
Note that the $N$-body model used here approximates UMa3/U1 as a collisionless system. The detailed stream properties will be affected by internal collisional processes, which in turn depend on the initial mass function, the fraction of binary stars, and the presence of dark stellar remnants \citep[see, e.g.][]{Spurzem2023}.

The bottom panel of Fig.~\ref{fig:selfgrav_lb} shows a close-up view of the present-day configuration of the simulated UMa3/U1 system, in Galactic coordinates. Because of the intrinsic faintness of the system, only a few individual stars are expected to trace the tidal tails outside the inner couple of arcminutes from the center of the system (the half-mass radius spans roughly $1'$ at $10\,\kpc$, the assumed distance of UMa3/U1; see \citetalias{Smith2024}).

We may use those stars to test the possibility that the velocity dispersion estimate of UMa3/U1 might be artificially enhanced by the presence of stars in the process of being stripped from the system. We find that the line-of-sight velocity dispersion using all stars within $1$, $2$, $3$, and $5$ half-mass radii varies by less than ${\sim}20\,\ms$ from its average value of ${\sim}60\,\ms$. In addition, the total line-of-sight velocity gradient across the galaxy is less than $25\,\ms/\mathrm{arcmin}$, so that stars $6'$ ahead or behind the  main body of the remnant differ by less than $300\,\ms$ on average, a value too small to be detectable with the line-of-sight velocities of \citetalias{Smith2024}. These results are not unexpected, given the extremely low escape velocity of the SG UMa3/U1 model, which (assuming the exponential profile of Eq.~\ref{eq:3DExp}) is only $v_\mathrm{esc} = (G M_\star / r_\star)^{1/2} \approx 215\,\ms$ at the center. For comparison, all of the likely UMa3/U1 members with available velocity estimates identified by \citetalias{Smith2024} lie within $4\, \Rh$ from the center of UMa3/U1.

The velocity dispersion estimate in the SG model thus seems to depend only weakly on the radial extent over which stars are collected, and certainly does not approach in any case the few km/s estimated by \citetalias{Smith2024} using 10 or 11 likely members (and neglecting binaries). We conclude that the inclusion of weakly bound stars, or of stars stirred by the Galactic tidal field, cannot explain such a high velocity dispersion estimate.

\subsection{Evaporation Timescales}
In addition to external tidal forces, internal collisional processes may alter the structure and bound mass of a stellar system. This is especially true for a system with as few stars as UMa3/U1, which may ``evaporate'' due to collisions between stars in less than a Hubble time. Assuming, for simplicity, that all stars have the same mass of ${\sim}0.25\,\Msol$, we find that  UMa3/U1 has approximately $N_\mathrm{h} \approx 2\,M_\star/\Msol \approx 32$ stars within the half-light radius.
For an isolated cluster, the relaxation time is related to the half-mass crossing time by (see e.g. Equation~(2-62) in \citealt{Spitzer1987book} and Equation~(8-1) in \citealt{BT87}):
\begin{equation}
T_\mathrm{rel} \approx \frac{N_\mathrm{h}}{8\,\ln N_\mathrm{h}} ~ T_\mathrm{cross} = (48 \pm 25) \,\Myr~ ,
\end{equation}
where $T_\mathrm{cross} \approx ( G M_\star/2 \rh^3)^{-1/2} = (42 \pm 22)\,\Myr$ for $M_\star = 16\, \Msol$ and $\rh=4$ pc.
The uncertainties quoted here are estimated by linear propagation of the measurement uncertainties. 

A rough estimate of the evaporation time scale is then given by \citep[see p.~491 in][]{BT87}:
\begin{equation}
 T_\mathrm{evap} \approx  136 ~T_\mathrm{rel} = (6.5\pm3.4)\,\Gyr ~.
 \label{eq:evaporation_time}
\end{equation}
This time scale is substantially longer than the time scale for full tidal disruption. Evaporation time estimates are shown as the dashed colored curves in Fig.~\ref{fig:SG_IC_systematics}. Collisional evaporation is thus unlikely to alter our main conclusion above: the short time to full tidal disruption clearly disfavors the suggestion that UMa3/U1 is a self-gravitating star cluster.

\section{Dark-matter-dominated Model}
\label{SecDM}
The simplest alternative to explain the long-term survival of UMa3/U1 in the Galactic tidal field is that UMa3/U1 is embedded in a dark matter subhalo, which protects the stellar component from tidal forces. The presence of dark matter would lead to a much increased velocity dispersion compared with the self-gravitating case studied in the previous section.

As a definite example, we shall adopt below the 10-member dispersion estimate of $\sigmalos=1.9\plus{1.4}\minus{1.1}\,\kms$ reported by \citetalias{Smith2024}, although we note again that this value may be revised once future observations enable a proper accounting of the effect of potential binary stars. We note as well that even a lower value of the velocity dispersion would qualify UMa3/U1 as a UFD, provided that it is substantially above the ${\sim} 50$ m/s expected from the SG model.

\subsection{Dynamical Mass Estimate}
\begin{figure}
\begin{center}
 \includegraphics[width=\columnwidth]{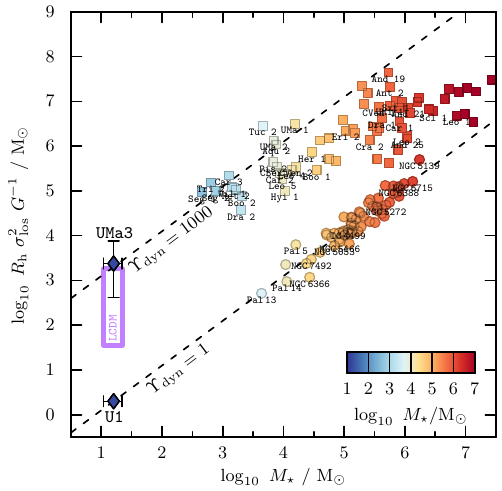}
\end{center}
 \caption{Dynamical mass, $\Rh \sigmalos^2 G^{-1}$, vs. stellar mass, $M_\star$, for Local Group dwarf galaxies (squares) and GCs (circles).
 The dashed diagonal lines indicate dynamical-to-stellar mass ratios of $\Upsilon_\mathrm{dyn}=1$ and $1000$, respectively (see Eq.~\ref{eq:dyn-to-stellar} for the definition). GCs are distributed with little scatter around the $\Upsilon_\mathrm{dyn}=1$ line. Dwarf galaxies, whose dynamics are dominated by dark matter, lie well above that line. Taking its measured velocity dispersion at face value, UMa-3 is located well within the LCDM prediction (the purple rectangle, see Sec.~\ref{sec:LCDMprediction}), whereas the SG model explored in Sec.~\ref{SecSG}, by construction, falls exactly on the  $\Upsilon_\mathrm{dyn}=1$ line. References for the data shown are listed in footnote~\ref{footnote:dsph_GC_data}.}
 \label{fig:mdyn_vs_L}
\end{figure}

The velocity dispersion adopted above would imply a dynamical-to-stellar mass ratio comparable to the most heavily dark-matter-dominated dwarfs known to date. We show this in Fig.~\ref{fig:mdyn_vs_L}, where we contrast UMa3/U1 with a compilation\footnote{The data shown for dwarf galaxies are as compiled in \citet{McConnachie2012} -- the version from 2021 January, with updated data for Antlia~2 \citep{Ji2021}, Bootes~2 \citep{Bruce2023}, Crater~2 \citep{Ji2021}, Tucana \citep{Taibi2020}, Tucana 2 \citep{Chiti2021}, And 19 \citep{Collins2020} and And 21 \citep{Collins2021}. For GCs, the data are taken from \citet{Harris1996} -- the version from 2010 December, with updated half-light radii and velocity dispersions for Pal-5 from \citet{Kuzma2015} and \citet{Gieles2021}; for NGC 2419 from \citet{Baumgardt2009}, and for Pal-14 from \citet{Hiker2006} and \citet{Jordi2009}.
\label{footnote:dsph_GC_data}} of dynamical and stellar masses of Local Group dwarf galaxies (squares) and GCs (circles) with measured kinematics.
The diagonal dashed curves correspond to constant dynamical-to-stellar mass ratios, approximated by
\begin{equation}
 \label{eq:dyn-to-stellar}
 \Upsilon_\mathrm{dyn} \equiv \frac{M_\mathrm{dyn}(<\rh)}{M_\star/2} \approx {8 \, \Rh \langle \sigmalos^{2} \rangle ~ G^{-1}}{M_\star^{-1}} ~.
\end{equation}
In the above equation, we estimate the dynamical mass $M_\mathrm{dyn}(<\rh)$ enclosed within the 3D half-light radius $\rh$ from the combined measurement of the line-of-sight velocity dispersion $\sigmalos$ and 2D half-light radius $\Rh$, with the coefficient\footnote{For dynamical masses enclosed within the 3D half-light radius, \citet{Wolf2010} and \citet{EPW18} give near-identical results. The derivation presented in the latter work, based on the projected virial theorem, guarantees that the results are independent of anisotropy in the velocity dispersion.\label{footnote:massest}} as in \citet{Wolf2010} and \citet{EPW18}.
The total stellar mass $M_\star$ is approximated assuming a stellar mass-to-light ratio of $M_\star/L_\mathrm{V}\approx1.6$ (as in \citealt{Woo2008} for galaxies with old stellar populations). 
Self-gravitating star clusters follow closely a line with constant dynamical-to-stellar mass ratio labeled $\Upsilon_\mathrm{dyn}=1$, whereas dwarfs are clearly offset to much higher values of the dynamical-to-stellar mass ratio.

\begin{figure}
\begin{center}
 \includegraphics[width=\columnwidth]{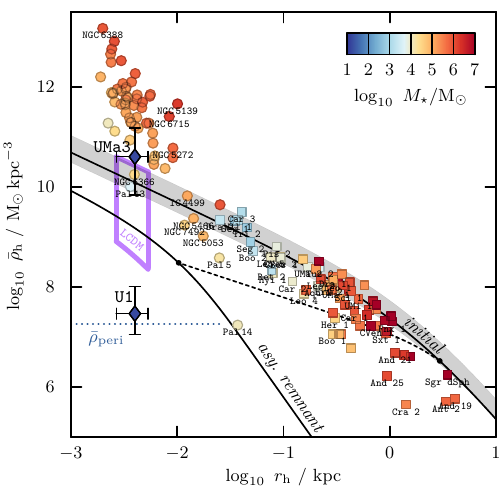}
\end{center}
 \caption{Mean density, $\bar \rho_\mathrm{h}$, enclosed within the 3D half-light radius, $\rh$, for Local Group dwarf galaxies (squares) and GCs (filled circles), compared with UMa3/U1. The diamond labeled "U1" shows the mean density expected if UMa3/U1 is a fully self-gravitating stellar system without dark matter. The upper diamond labelled ``UMa3''' corresponds to adopting the measured velocity dispersion of $1.9\,\kms$. A gray band shows the mean enclosed densities as a function of radius for LCDM (NFW) halos considered sufficiently massive to allow stars to form, taking into account the expected scatter in concentration (see the text for details). An example halo (labeled ``initial'') with a virial mass of $M_\mathrm{cr}^{z=2} \approx 9.5\times10^8\, M_\odot$, corresponding to the $z=2$ hydrogen-cooling critical mass with average concentration, is shown in black. The dashed curve illustrates the ``tidal track'' tracing the evolution of the characteristic density $\bar \rho_\mathrm{mx} = \bar \rho(<\rmx)$ and size $\rmx$ of an NFW halo as it is stripped by tides, with the black circles highlighting the initial and asymptotic values. The  lower black curve corresponds to the asymptotic remnant of the ``initial'' model placed on the UMa3/U1 orbit.}
 \label{fig:rho_vs_rh}
\end{figure}

The blue diamond at the bottom left corner shows where UMa3/U1 would be located if it was a faint GC akin to the SG model explored earlier (in which case it would be called ``UNIONS~1'', or U1, for short). The other blue diamond labeled ``UMa3'' corresponds to adopting the 10-star dispersion $\sigmalos=1.9\plus{1.4}\minus{1.1}\,\kms$, resulting in a dynamical-to-stellar mass ratio of $\Upsilon_\mathrm{dyn}=1.2\plus{2.8}\minus{1.0}\,\times10^3$. In this case, UMa3 sits comfortably close to the dwarf galaxy trend, extrapolated to extremely low stellar masses (i.e., extremely faint luminosities).

Assuming hereafter that UMa3/U1 is a dwarf galaxy (in which case it would simply be called ``Ursa Major III'' or UMa3, for short), we examine next whether it would be expected to survive the strong Galactic tidal field. We begin by comparing in Fig.~\ref{fig:rho_vs_rh} the mean density of the system within its half-mass radius, ${\bar \rho}_\mathrm{h}$, with ${\bar \rho}_\mathrm{peri}$, the mean density of the Galaxy at the pericenter of the orbit, shown by a horizontal dotted line segment. As discussed earlier, the ``U1'' symbol indicates that UMa3/U1's density would be comparable to  ${\bar \rho}_\mathrm{peri}$ (and thus doomed to rapid tidal disruption) if it was a self-gravitating cluster.

We estimate the density of UMa3 by computing the dynamical mass enclosed within its 3D half-light radius using (see footnote~\ref{footnote:massest})
\begin{equation}
 M_\mathrm{dyn}(<\rh) \approx  4~ \Rh  \langle \sigmalos^2 \rangle  G^{-1} = \left(1.0\plus{2.1}\minus{0.8}\right) \times 10^4 \,\Msol~,
 \label{eq:dynmass}
\end{equation}
and dividing by the volume of a sphere of radius $\rh=(4/3)\,\Rh$.  If $\sigmalos=1.9\plus{1.4}\minus{1.1}$ km/s, this yields
\begin{equation}
\label{eq:rho_dyn}
 \bar \rho_\mathrm{h}=\left(4\plus{11}\minus{3} \right)\times 10^{10}\, \Msol\,\kpc^{-3} ~,
\end{equation}
which is roughly ${\sim}1000$ times higher than the density of the SG model, and thus safe from tidal disruption.
As shown by the blue diamond labeled ``UMa3'' in Fig.~\ref{fig:rho_vs_rh},  the mean density of UMa3 would in that case be comparable to that of some GCs, many of which are known to orbit the Galaxy on orbits with pericenters as small as or smaller than $13$ kpc.

The mean density computed above would make UMa3 not only the faintest and smallest, but also the densest UFD ever detected, with important implications for both the mass of the CDM halo inhabited by UMa3 and for alternative models of dark matter. We address these issues next.

\subsection{LCDM Expectations}
\label{sec:LCDMprediction}
In LCDM, galaxies form deep within the potential wells of dark matter halos \citep{WhiteRees1978}. The ability of hydrogen gas to cool efficiently in the presence of the cosmic UV background is expected to impose a minimum \emph{critical} halo mass below which LCDM halos are not expected to be able to host luminous galaxies.

\citet{Benitez-Llambay2020} argue that, after reionization, the critical virial mass needed to enable star formation to proceed evolves with redshift $z$ roughly as
\begin{equation}
  M_\mathrm{cr}^z \approx (10^{10}\Msol) \left[T / (3.2\times10^4\,\mathrm{K})\right]^{3/2} (1+z)^{-3/2}~,
  \label{EqMcrit}
\end{equation}
using a virial temperature of $T= 2 \times 10^4\,\mathrm{K}$. At redshift $z=0$, the resulting critical mass equals $M_\mathrm{cr}^{z=0} \approx 4.9 \times  10^9\,\Msol$.

At earlier redshifts, the critical mass was somewhat lower. The best-fitting isochrone for U1/UMa3 corresponds to a stellar age of $\gtrsim 11\,\Gyr$ \citepalias{Smith2024}.
Assuming \citet{Planck2020} cosmological parameters, this corresponds roughly to a redshift of $z \sim 2$, which we shall adopt in what follows. 

The cosmological simulations analyzed by \citet{Pereira-Wilson23} confirm that stars first form in halos exceeding the critical mass given by Eq.~\ref{EqMcrit}. We use their Figure~7  to estimate a range of halo masses for the potential progenitor halo of UMa3 (up to $0.5$ dex above $M_\mathrm{cr}$) . Combining this with the $z=2$ average NFW concentration ($\pm0.15\,\dex$ scatter) from \citet{Ludlow2016}, we obtain a range of mass profiles for UMa3, which we show as a gray band in Fig.~\ref{fig:rho_vs_rh}. The mean density profile of a halo with virial mass $M_\mathrm{cr}^{z=2} \approx 9.5 \times 10^8 \,\Msol$ is shown as a solid black curve within the gray band.

Fig.~\ref{fig:rho_vs_rh} shows that UMa3's mass density is (for the assumed $\sigmalos$) comfortably within the range expected for a galaxy inhabiting a cuspy NFW halo with mass close to critical. Less massive NFW halos are less dense at all radii, and would therefore have difficulty matching UMa3's estimated density. For example, an LCDM ``minihalo'' with virial mass $10^6\, \Msol$ at $z=0$ would have a mean density of $\approx 1.2 \times 10^9\, \Msol\,\kpc^{-3}$ at $r\sim 4$ pc, well below UMa3.

The high dark matter density estimated for UMa3 thus disfavors the possibility that it may have formed at the center of a minihalo and supports the view that luminous galaxies, no matter how faint, only form in halos near or above the critical virial mass of \citet{Benitez-Llambay2020}.

If UMa3 is indeed a ``micro galaxy'' \citepalias{EP20} at the center of a cuspy NFW halo, we may use the results of \citetalias{EN21} and \citet{ENIP2022} to predict its evolution under the influence of the Galactic tidal field\footnote{We compare the tidal evolution of UMa3 as predicted by the \citetalias{EN21} model against controlled $N$-body simulations in Appendix~\ref{Appendix_Nbody}.}. The main result of their study is that tides gradually strip a halo, approaching an asymptotic state where the  characteristic density of the bound remnant equals roughly $16 \times {\bar\rho}_\mathrm{peri}$. The subhalo characteristic density evolves following a well-defined ``tidal track'' \citep{Penarrubia2008} until the asymptotic characteristic density has been reached. The bound remnant is well approximated by an ``exponentially truncated'' NFW profile \citep[see][for details]{EN21}.

The initial density profile of a halo of critical mass $M_\mathrm{cr}^{z=2}$, is shown by the solid black curve in Fig.~\ref{fig:rho_vs_rh}, labeled ``initial". This halo has a circular velocity that peaks at $\Vmx = 23\,\kms$ at a radius $\rmx = 3.0\,\kpc$. Its characteristic mean density at $\rmx$ equals $\bar \rho_\mathrm{mx} = 3.3\times10^6\,\Msol\,\kpc^{-3}$, shown as a black circle in Fig.~\ref{fig:rho_vs_rh}.
The tidal track is shown by the dashed black curve, which stops once the asymptotic remnant density of $\bar \rho_\mathrm{mx} \approx 16 \times {\bar\rho}_\mathrm{peri}$ has been reached. An exponentially truncated NFW profile (the black curve labeled ``asy. remnant'') illustrates the final density profile of that halo on this orbit.

We emphasize that the asymptotic remnant properties depend on those of the initial NFW halo adopted. Choosing an initial halo with a higher characteristic initial density would lead to an asymptotic remnant whose mean density, at $r=4\,\pc$, is higher. It is clear from Fig.~\ref{fig:rho_vs_rh} that although Galactic tides are expected to reduce the central dark matter density of a cuspy halo, a well-characterized bound remnant is predicted to survive, whose density at radii as small as $4\,\pc$ may be as high as ${\sim} 10^{10}\, \Msol/$kpc$^3$.

The full range of dynamical masses and densities expected at that radius in LCDM (varying halo mass and concentration) is shown by the purple curves labeled ``LCDM'' in Fig.~\ref{fig:mdyn_vs_L} and Fig.~\ref{fig:rho_vs_rh}, respectively. The corresponding range of velocity dispersions is  $0.16 \lesssim \sigmalos / \kms \lesssim 1.5$, which  is compatible with the current 10-member dispersion estimate of $\sigmalos=1.9\plus{1.4}\minus{1.1}\,\kms$ \citepalias{Smith2024}. 
We conclude that UMa3's properties are consistent with those expected from a micro galaxy deeply embedded at the center of a fairly massive, cuspy LCDM halo.

\subsection{Consequences for Alternative Dark Matter Models}
Although consistent with LCDM, the extreme properties of UMa3 inferred assuming that the 10-member velocity dispersion measurement holds ($\sigmalos=1.9\plus{1.4}\minus{1.1}\,\kms$) would be difficult to reconcile with alternative dark matter models that predict lower dark matter densities.

We begin by discussing UFDs in FDM models, where the density of a dark-matter-dominated UFD is thought to reflect the central density of the ``solitonic core'' that forms at the center of a halo made up of ultralight particles \citep{Schive2014,Schive2014b,Safarzadeh2020}. Cosmological simulations of FDM halo formation find that the central density of the core, at redshift $z=0$, is given by (using Equation~3 and Equation~7 in \citealt{Schive2014b})
\begin{equation}
  \rho_\mathrm{c0} \approx 2.9 \times 10^{6}\,\frac{\Msol}{\kpc^{3}}~ \left(\frac{m_\psi}{10^{-22}\,\mathrm{eV}/\mathrm{c}^2} \right)^2 \, \left( \frac{M_\mathrm{halo}}{10^9\,\Msol} \right)^{4/3}  ~,
\label{EqFDM}
\end{equation}
where $m_\psi$ is the mass of the ultralight particle, and $M_\mathrm{halo}$ is a measure of the halo virial mass\footnote{The definitions of the halo virial mass $M_\mathrm{halo}$ and virial radius in \citet{Schive2014b} are slightly different from the ones we use throughout this work (see footnote~\ref{footnote:virial}): at redshift $z=0$, the mean density enclosed within the virial radius equals $\approx350$ times the mean density of the Universe.}.

Attempts to fit the density profile of the ``classical'' dwarf spheroidal satellites of the Milky Way (like Fornax or Sculptor) with a solitonic core yield upper limits for $m_\psi$ of order $10^{-22}\,\mathrm{eV}/\mathrm{c}^2$ \citep{Marsh2015,GonzalezMorales2017}. This is mainly because the mean densities of Fornax and Sculptor are of order $10^7$--$10^8\, \Msol\,\kpc^{-3}$, consistent with Eq.~\ref{EqFDM} for $M_\mathrm{halo} \sim 10^{10}\,\Msol$ and $m_\psi \sim 10^{-22}\,\mathrm{eV}/\mathrm{c}^2$.

Achieving a density as high as that estimated for UMa3 (i.e., ${\sim} 4\times 10^{10}\, \Msol\,\kpc^{-3}$, see Eq.~\ref{eq:rho_dyn}) would require the ultra-light particle mass to be as large as $m_\psi \sim 3\times 10^{-21}\,\mathrm{eV}/\mathrm{c}^2$ (or a halo mass as large as $M_\mathrm{halo} \sim10^{12}\, \Msol$, which seems rather unlikely). This choice, however, would deny the main motivation for FDM: that the presumed kpc-scale cores in dwarf galaxies suggested by some studies reflect the de Broglie wavelength of the ultralight particle \citep{Goodman2000,Hu2000,Hui2017,Ferreira2021}. Reconciling FDM with the high densities of observed UFDs has already been recognized as a difficult challenge to traditional FDM models \citep{Burkert2020,Safarzadeh2020}, a challenge that would become much more severe if the high density of UMa3 is confirmed by future velocity dispersion measurements. 

Finally, the estimated high density of UMa3 is also difficult to accommodate with the kpc-size cores expected in self-interacting dark matter models, at least for interaction cross sections of order $1\,\mathrm{cm}^2\mathrm{g}^{-1}$ \citep[see; e.g.,][and references therein]{Tulin2018}. For this choice, collisions between particles erase the central cusp of a dark matter halo, creating a core with constant density that, on the scale of dwarf galaxies, does not exceed a few times $10^8\, \Msol\,\kpc^{-3}$ \citep{Zavala2013, Vogelsberger2014}. 
Matching the high density of UMa3 (and, indeed, of other UFDs) would require gravothermal ``core collapse'' to occur, 
raising the innermost dark matter densities to values as large as those predicted for cuspy LCDM halos, and observed in UFDs \citep[see, e.g.][]{Hayashi2021, Silverman2023}.

The timescale for core collapse, however, likely exceeds the age of the Universe for SIDM models with velocity-independent interaction cross sections (\citealt{Zeng2022}; though some authors argue that the core collapse timescales may be shortened considerably by tidal effects, see \citealt{Nishikawa2020}). Further work is clearly needed to reconcile SIDM models with the high densities of UFDs in general, and of UMa3 in particular, should the current density estimate prove robust to binary stars.

\subsection{Annihilation Signals}
The remarkably high density of $\bar \rho_\mathrm{h} \sim 4\times10^{10} \,\Msol\kpc^{-3}$ (Eq.~\ref{eq:rho_dyn}) combined with the heliocentric distance of only $(10\pm1)\,\kpc$ \citepalias{Smith2024} render UMa3/U1 an interesting target for the study of potential signals of dark matter self-annihilation.
The astrophysical component to the annihilation signal for velocity-independent annihiliation may be expressed through the $J$-factor \citep[see, e.g.][]{Walker2011_Jfact}, which for stellar tracers embedded in an NFW subhalo can be estimated from \citep[see Equation~13 in][]{Evans2016}
\begin{equation}
 J \equiv\!\iint\! \diff l \diff \Omega\, \rho_\mathrm{NFW}^2 \approx \frac{25}{8 G^2} \frac{\sigmalos^4 \theta}{D \Rh^2} \sim 10^{21} \, \frac{\mathrm{GeV}^2/\mathrm{c}^4}{\mathrm{cm}^5}~.
 \label{eq:J-fact}
\end{equation}
In the above equation, the integral is performed along the line of sight $l$ over a solid angle $\Delta\Omega$, and $D$, $\Rh$, and $\sigmalos$ are the  heliocentric distance, projected half-light radius, and line-of-sight velocity dispersion of UMa3/U1, respectively. The angle $\theta$ limits the solid angle over which the integral is computed, and is chosen here to match the Fermi Large Area Telescope resolution at GeV scales, $\theta=0^\circ.5$ \citep{Ackermann2014}. The $J$-factor of ${\sim} 10^{21}\,\mathrm{GeV}^2/\mathrm{c}^4/\mathrm{cm}^5$ computed in Eq.~\ref{eq:J-fact} takes the measured properties of UMa3/U1 at face value. Monte-Carlo sampling of measurement uncertainties yields a $16^\mathrm{th}$--$84^\mathrm{th}$ percentile range of $10^{19} \lesssim J / (\mathrm{GeV}^2/\mathrm{c}^4/\mathrm{cm}^5) \lesssim 10^{22}$ for the underlying distribution. Even when taking these large uncertainties into account, UMa3/U1 would be one of the ``brightest'' satellites of the Milky Way for potential annihilation signals, matching or exceeding the expected signal of all other known Milky Way dwarf spheroidal and ultrafaint satellites (see Table~A2 in \citealt{PaceStrigari2019}).

\section{Summary and Conclusions}
\label{SecConc}

Ursa Major III/UNIONS 1 (UMa3/U1) is a recently discovered satellite of the Milky Way whose extreme properties offer unique insights into the formation process of some of the faintest objects in the Universe. It is by far the faintest satellite ever discovered: at $M_\mathrm{V}\approx+2.2$, it is ${\sim}10$ times fainter than the faintest confirmed ultrafaint dwarfs, and ${\sim}10$ times fainter than the least luminous globular cluster. It is also as small as some of the most compact GCs, with a projected half-mass radius of only ${\sim} 3$ pc.

Taken at face value, the line-of-sight velocity dispersion computed from the radial velocities of 10 or 11 likely members suggests the presence of dark matter (which would confirm that UMa3/U1 is indeed a dwarf galaxy). However, the lack of repeat velocity measurements and the strong dependence of the measured dispersion on the inclusion of two specific member stars leave open the possibility that UMa3/U1 is actually a self-gravitating faint cluster of stars.

The main conclusions of our work are summarized below.

(1) The orbit of the system around the Milky Way is well characterized, with a pericentric distance of only $\approx13\,\kpc$, an apocentric distance of $\approx30\,\kpc$, and a radial orbital time of $\approx 0.4\,\Gyr$. We use N-body simulations to show that, taken its observed size and stellar mass at face value, UMa3/U1 cannot survive for much longer than a single orbit if self-gravitating. Either UMa3/U1 is a GC remnant observed at a remarkably fine-tuned point in time, or it is indeed a galaxy that has survived tidal disruption because of the stabilizing effect of dark matter.

(2) The simulations do rule out the possibility that ``tidal stirring'' may have enhanced the observed $\sigmalos$ to values as high as a few km/s. The models predict that, if self-gravitating, UMa3/U1 should be surrounded by a stream of tidally stripped stars, that should be searched for with high priority.

We conclude that (1) and (2) strongly support the view that UMa3/U1 is a dark matter-dominated ``micro galaxy'', indeed the faintest, or ``darkest'', galaxy ever discovered.

(3) If the $\sigmalos=1.9\plus{1.4}\minus{1.1}\,\kms$ estimated using the 10 most-likely member stars is correct, then the mean dark matter density for  UMa3/U1 would be $\approx 4\times 10^{10}\,\Msol\,\kpc^{-3}$ within its 3D half-light radius of ${\sim} 4\,\pc$. This makes UMa3/U1 the densest ultra-faint galaxy known, a result that suggests that UMa3/U1 formed at the very center of a fairly massive (${\sim} 10^9$-$10^{10}\, \Msol$) cuspy cold dark matter halo, whose innermost regions should be able to survive the strong tidal field of the Galaxy for a Hubble time. This in turn implies a relatively high halo mass threshold for luminous galaxy formation in LCDM, as advocated by the ``critical'' mass model of \citet{Benitez-Llambay2020}.

(4) If confirmed, the high dark matter density estimated for UMa3/U1 would have strong implications for alternative dark matter models. In the case of ``fuzzy dark matter'' (FDM), identifying it with the central density of the ``solitonic core'' yields an estimate for the mass of the ultra-light particle of ${\sim} 3\times 10^{-21}\,\mathrm{eV}$, more than one order of magnitude higher than the $10^{-22}\,\mathrm{eV}$ upper limits inferred in previous work \citep{Marsh2015,GonzalezMorales2017}. This inconsistency casts doubt on the ground motivation for FDM models, which, if confirmed, would require a re-evaluation of the model.

(5) Such a high dark matter density would also place strong constraints on SIDM models, where the dense halo cusps are eroded by collisional effects, leading to a substantial reduction of the dark matter central densities compared to LCDM. In the context of SIDM, UMa3/U1's central density 
can only be reproduced in systems that have undergone gravothermal ``core collapse'', placing stringent constraints on the allowed values of the collisional cross section.

Our models show that accurate and extremely precise dispersion estimates are crucial to differentiating between UMa3/U1 being a dark-matter-dominated ``micro galaxy'', or an extremely faint self-gravitating star cluster. If the current estimate of UMa3's dynamical density is confirmed by  future observations,  it would not only confirm UMa3 as the ``darkest'' galaxy discovered to date, but it would also highlight that the predictions of LCDM seem to hold down to the faintest end of the galaxy luminosity function.

\section*{Acknowledgments}
The authors would like to thank W. Cerny, T. S. Li and M. G. Walker for their helpful comments on the draft. RE acknowledges support from the European Research Council (ERC) under the European Union's Horizon 2020 research and innovation program (grant agreement No. 834148), and from the National Science Foundation (NSF) grant AST-2206046. 

\bigskip

\appendix
\section{Tidal Mass Loss on Different Orbits}
\label{Appendix_orbits}
The tidal evolution and potential disruption of UMa3/U1 depend on its orbit. In Sec.~\ref{SecSG}, we use $N$-body simulations to study the tidal disruption of UMa3/U1 in the \citetalias{EP20} Milky Way potential, taking the observed position of UMa3/U1 at face value. In the following, we will extend the analysis to include a second potential, and account for observational uncertainties in the present position and velocity of UMa3/U1.

\begin{figure}
\begin{center}
 \includegraphics[width=\columnwidth]{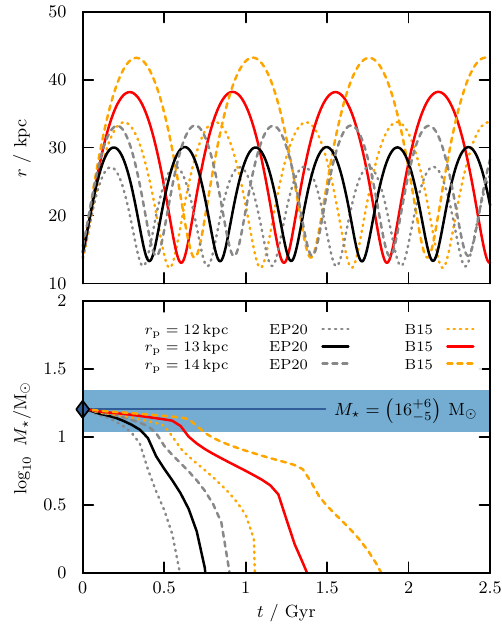}
\end{center}
 \caption{Top panel: galactocentric distance of UMa3/U1 as a function of time, obtained by forward integrating orbits in the  \citetalias{EP20} (black curves) and \citep[][red curves]{Bovy2015} potentials. The solid curves take the current position and velocity of UMa3/U1 at face value. The dotted and dashed lines correspond to orbits representative of the $16^\mathrm{th}$ and  $84^\mathrm{th}$ percentiles of the underlying distribution of measurement uncertainties.
 Bottom panel: the evolution of the bound mass of self-gravitating (SG) $N$-body models evolved on the orbits shown above. In all cases, the SG models disrupt within a few pericenter passages.}
 \label{fig:orbit_exploration}
\end{figure}

The axisymmetric \citetalias{EP20} Milky Way potential has a circular velocity of $\Vc(R_\odot) = 240\,\kms$ at the solar radius. As a second example potential, we use the \citet{Bovy2015} model with $\Vc(R_\odot) = 220\,\kms$. These two potential models nicely bracket the rotation curve measured by \citet{Eilers2019}. We integrate point-mass orbits in these two potentials. The initial conditions are obtained by drawing Monte Carlo samples from the distribution of measured distance, radial velocity and proper motions of UMa3/U1, with uncertainties as listed in \citetalias{Smith2024}. 
For the \citetalias{EP20} potential, we obtain pericenter- and apocenter estimates of $\rperi = (13\pm1)\,\kpc$ and $\rapo = (30\plus{4}\minus{3})\,\kpc$. For the case of the \citet{Bovy2015} potential, our pericenter estimate is virtually identical, whereas for the apocenter, we find $\rapo = (38\plus{6}\minus{5})\,\kpc$. The uncertainties quoted correspond to the $16^\mathrm{th}$ and  $84^\mathrm{th}$ percentiles of the underlying distribution.

The top panel of Figure~\ref{fig:orbit_exploration} shows the galactocentric distance of UMa3/U1 as a function of time for three orbits in the \citetalias{EP20} potential, and three orbits in the \citet{Bovy2015} potential. The solid curves show the orbit obtained by taking the measured position and velocity of UMa3/U1 at face value, while the dotted and dashed curves show orbits representative for the $16^\mathrm{th}$ and  $84^\mathrm{th}$ percentile of the underlying distribution. 

For each of the orbits shown in the top panel of Fig.~\ref{fig:orbit_exploration}, we evolve an $N$-body model of UMa3/U1 forward in time, using the same setup as in Sec.~\ref{SecSG}. The initial stellar mass and half-light radius are chosen to match the currently observed values of $M_\star = 16\,\Msol$ and $\Rh = 3\,\pc$. We show the evolution of the bound mass for each of these $N$-body models in the bottom panel of Figure~\ref{fig:orbit_exploration}. All models disrupt within the next few pericenter passages, showing that the results of Sec.~\ref{SecSG} are fairly insensitive to the choice of the Milky Way potential, and fairly robust within the estimated uncertainties of the observed position and velocity of UMa3/U1.

\begin{figure}
\begin{center}
 \includegraphics[width=\columnwidth]{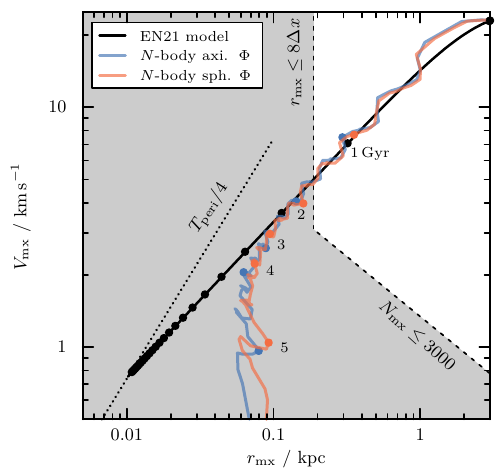}
\end{center}
 \caption{Tidal evolution of characteristic size $\rmx$ and velocity $\Vmx$ for an NFW subhalo on the orbit of UMa3/U1. The initial values ($r\mxzero = 3\,\kpc$, $V\mxzero = 23\,\kms$) are identical to the subhalo model discussed in Sec.~\ref{SecDM}. A black solid curve shows the evolution computed using the \citetalias{EN21} model, while the blue and red curves correspond to $N$-body models evolved in an axisymmetric and a spherical host potential, respectively. The gray shaded area shows the parameter space where the $N$-body models are likely affected by insufficient particle numbers $(N_\mathrm{mx} \leq 3000)$ and/or spatial resolution ($\rmx \leq 8 \Delta x$). Both $N$-body models disrupt \textit{artificially} after $\approx 5\,\Gyr$, whereas the \citetalias{EN21} model predicts an evolution toward a stable remnant state.}
 \label{fig:nbody_vs_EN21}
\end{figure}

\section{Supplementary Dark-matter-only \textit{N}-body Simulations}
\label{Appendix_Nbody}
In Sec.~\ref{SecDM}, we discuss the observable consequences for a scenario where UMa3/U1 is embedded in a dark matter subhalo. Our analysis makes use of the empirical findings of \citetalias{EN21}, which suggest that tidally stripped NFW halos converge towards a stable asymptotic remnant state, where the characteristic density of the subhalo is determined by the mean density of the host at pericenter. The model of \citetalias{EN21} assumes a spherical, isothermal host potential. The combined mass distribution of the halo, the bulge and the disk in the inner regions of the Milky Way is not spherical, however, but more appropriately approximated by an axisymmetric model. Using $N$-body simulations, we now aim to test to what extent the tidal evolution of a dark matter subhalo on the orbit of UMa3/U1 depends on the geometry of the underlying potential. 

We place two NFW $N$-body subhalo models on orbits with pericenter and apocenter matching those of UMa3/U1, and evolve these models in (1) the axisymmetric \citetalias{EP20} potential and (2) a spherical isothermal potential with a constant circular velocity of $V_\mathrm{c} = 240\,\kms$. The $N$-body subhalo models are chosen to have initial properties identical to those of the $z=2$ hydrogen cooling limit (HCL) halo of Fig.~\ref{fig:rho_vs_rh}, and $N=10^7$ particles. We generate the models using the same code as in Sec.~\ref{SecSG} (see footnote~\ref{footnote:nbopy}), and evolve the models using the particle mesh code \textsc{superbox} \citep{Fellhauer2000}, adopting a spatial resolution of $\Delta x \approx 20\,\pc$ for the highest-resolving grid.

The results of the experiment are shown in Fig.~\ref{fig:nbody_vs_EN21}, where we confront the time evolution of the subhalo structural parameters $\rmx$, $\Vmx$ measured in the two $N$-body models against those computed using the \citetalias{EN21} model. 
For the first ${\sim} 2\,\Gyr$ of tidal evolution, both $N$-body models are in good agreement with \citetalias{EN21}.
The gray shaded area in Fig.~\ref{fig:nbody_vs_EN21} shows the region of parameter space where the number of $N$-body particles ($N_\mathrm{mx} \leq 3000$) and/or the spatial resolution of the simulation ($\rmx \leq 8\Delta x$) are insufficient to reliably model the tidal evolution of the subhalo (see Appendix~A in \citetalias{EN21} for a convergence study). Indeed, in the gray shaded region, the evolution of the $N$-body models diverges from the \citetalias{EN21} track, and both models \textit{artificially} disrupt after ${\sim} 5\,\Gyr$ of evolution. The \citetalias{EN21} model instead predicts an evolution toward a stable remnant state. The filled circles along the evolutionary track are spaced by $1\,\Gyr$, and show that the remnant state is approached asymptotically, with tidal evolution gradually slowing down as the remnant state is approached. 

Remarkably, the tidal evolution of the $N$-body model evolved in the axisymmetric potential is near identical to the one evolved in the spherical potential. For the orbit of UMa3/U1, the detailed shape of the potential hence seems to have negligible impact on the tidal evolution of a dark matter subhalo.

\bigskip

\bibliographystyle{aasjournal}
\bibliography{unions1}{}

\begin{thebibliography}{}
\expandafter\ifx\csname natexlab\endcsname\relax\def\natexlab#1{#1}\fi
\providecommand{\url}[1]{\href{#1}{#1}}
\providecommand{\dodoi}[1]{doi:~\href{http://doi.org/#1}{\nolinkurl{#1}}}
\providecommand{\doeprint}[1]{\href{http://ascl.net/#1}{\nolinkurl{http://ascl.net/#1}}}
\providecommand{\doarXiv}[1]{\href{https://arxiv.org/abs/#1}{\nolinkurl{https://arxiv.org/abs/#1}}}

\bibitem[{{Ackermann} {et~al.}(2014){Ackermann}, {Albert}, {Anderson},
  {Baldini}, {Ballet}, {Barbiellini}, {Bastieri}, {Bechtol}, {Bellazzini},
  {Bissaldi}, {Bloom}, {Bonamente}, {Bouvier}, {Brandt}, {Bregeon}, {Brigida},
  {Bruel}, {Buehler}, {Buson}, {Caliandro}, {Cameron}, {Caragiulo}, {Caraveo},
  {Cecchi}, {Charles}, {Chekhtman}, {Chiang}, {Ciprini}, {Claus},
  {Cohen-Tanugi}, {Conrad}, {D'Ammando}, {de Angelis}, {Dermer}, {Digel}, {do
  Couto e Silva}, {Drell}, {Drlica-Wagner}, {Essig}, {Favuzzi}, {Ferrara},
  {Franckowiak}, {Fukazawa}, {Funk}, {Fusco}, {Gargano}, {Gasparrini},
  {Giglietto}, {Giroletti}, {Godfrey}, {Gomez-Vargas}, {Grenier}, {Guiriec},
  {Gustafsson}, {Hayashida}, {Hays}, {Hewitt}, {Hughes}, {Jogler}, {Kamae},
  {Kn{\"o}dlseder}, {Kocevski}, {Kuss}, {Larsson}, {Latronico}, {Llena Garde},
  {Longo}, {Loparco}, {Lovellette}, {Lubrano}, {Martinez}, {Mayer},
  {Mazziotta}, {Michelson}, {Mitthumsiri}, {Mizuno}, {Moiseev}, {Monzani},
  {Morselli}, {Moskalenko}, {Murgia}, {Nemmen}, {Nuss}, {Ohsugi}, {Orlando},
  {Ormes}, {Perkins}, {Piron}, {Pivato}, {Porter}, {Rain{\`o}}, {Rando},
  {Razzano}, {Razzaque}, {Reimer}, {Reimer}, {Ritz}, {S{\'a}nchez-Conde},
  {Sehgal}, {Sgr{\`o}}, {Siskind}, {Spinelli}, {Strigari}, {Suson}, {Tajima},
  {Takahashi}, {Thayer}, {Tibaldo}, {Tinivella}, {Torres}, {Uchiyama}, {Usher},
  {Vandenbroucke}, {Vianello}, {Vitale}, {Werner}, {Winer}, {Wood}, {Wood},
  {Zaharijas}, {Zimmer}, \& {Fermi-LAT Collaboration}}]{Ackermann2014}
{Ackermann}, M., {Albert}, A., {Anderson}, B., {et~al.} 2014, \prd, 89, 042001,
  \dodoi{10.1103/PhysRevD.89.042001}

\bibitem[{{Amorisco}(2021)}]{Amorisco2021}
{Amorisco}, N.~C. 2021, arXiv e-prints, arXiv:2111.01148.
\newblock \doarXiv{2111.01148}

\bibitem[{{Amorisco} \& {Evans}(2012)}]{Amorisco2012}
{Amorisco}, N.~C., \& {Evans}, N.~W. 2012, \mnras, 419, 184,
  \dodoi{10.1111/j.1365-2966.2011.19684.x}

\bibitem[{{Balberg} {et~al.}(2002){Balberg}, {Shapiro}, \&
  {Inagaki}}]{Balberg2002}
{Balberg}, S., {Shapiro}, S.~L., \& {Inagaki}, S. 2002, \apj, 568, 475,
  \dodoi{10.1086/339038}

\bibitem[{{Balbinot} {et~al.}(2013){Balbinot}, {Santiago}, {da Costa}, {Maia},
  {Majewski}, {Nidever}, {Rocha-Pinto}, {Thomas}, {Wechsler}, \&
  {Yanny}}]{Balbinot2013}
{Balbinot}, E., {Santiago}, B.~X., {da Costa}, L., {et~al.} 2013, \apj, 767,
  101, \dodoi{10.1088/0004-637X/767/2/101}

\bibitem[{{Battaglia} \& {Nipoti}(2022)}]{BattagliaNipoti2022Review}
{Battaglia}, G., \& {Nipoti}, C. 2022, Nature Astronomy, 6, 659,
  \dodoi{10.1038/s41550-022-01638-7}

\bibitem[{{Baumgardt} {et~al.}(2009){Baumgardt}, {C{\^o}t{\'e}}, {Hilker},
  {Rejkuba}, {Mieske}, {Djorgovski}, \& {Stetson}}]{Baumgardt2009}
{Baumgardt}, H., {C{\^o}t{\'e}}, P., {Hilker}, M., {et~al.} 2009, \mnras, 396,
  2051, \dodoi{10.1111/j.1365-2966.2009.14932.x}

\bibitem[{{Belokurov} {et~al.}(2009){Belokurov}, {Walker}, {Evans}, {Gilmore},
  {Irwin}, {Mateo}, {Mayer}, {Olszewski}, {Bechtold}, \&
  {Pickering}}]{Belokurov2009}
{Belokurov}, V., {Walker}, M.~G., {Evans}, N.~W., {et~al.} 2009, \mnras, 397,
  1748, \dodoi{10.1111/j.1365-2966.2009.15106.x}

\bibitem[{{Benitez-Llambay} \& {Frenk}(2020)}]{Benitez-Llambay2020}
{Benitez-Llambay}, A., \& {Frenk}, C. 2020, \mnras, 498, 4887,
  \dodoi{10.1093/mnras/staa2698}

\bibitem[{{Binney} \& {Tremaine}(1987)}]{BT87}
{Binney}, J., \& {Tremaine}, S. 1987, {Galactic dynamics}

\bibitem[{{Bode} {et~al.}(2001){Bode}, {Ostriker}, \& {Turok}}]{Bode2001}
{Bode}, P., {Ostriker}, J.~P., \& {Turok}, N. 2001, \apj, 556, 93,
  \dodoi{10.1086/321541}

\bibitem[{{Bovy}(2015)}]{Bovy2015}
{Bovy}, J. 2015, \apjs, 216, 29, \dodoi{10.1088/0067-0049/216/2/29}

\bibitem[{{Bruce} {et~al.}(2023){Bruce}, {Li}, {Pace}, {Heiger}, {Song}, \&
  {Simon}}]{Bruce2023}
{Bruce}, J., {Li}, T.~S., {Pace}, A.~B., {et~al.} 2023, \apj, 950, 167,
  \dodoi{10.3847/1538-4357/acc943}

\bibitem[{{Bullock} \& {Boylan-Kolchin}(2017)}]{Bullock2017_Review}
{Bullock}, J.~S., \& {Boylan-Kolchin}, M. 2017, \araa, 55, 343,
  \dodoi{10.1146/annurev-astro-091916-055313}

\bibitem[{{Burkert}(2020)}]{Burkert2020}
{Burkert}, A. 2020, \apj, 904, 161, \dodoi{10.3847/1538-4357/abb242}

\bibitem[{{Cerny} {et~al.}(2023{\natexlab{a}}){Cerny},
  {Mart{\'\i}nez-V{\'a}zquez}, {Drlica-Wagner}, {Pace}, {Mutlu-Pakdil}, {Li},
  {Riley}, {Crnojevi{\'c}}, {Bom}, {Carballo-Bello}, {Carlin}, {Chiti}, {Choi},
  {Collins}, {Darragh-Ford}, {Ferguson}, {Geha}, {Mart{\'\i}nez-Delgado},
  {Massana}, {Mau}, {Medina}, {Mu{\~n}oz}, {Nadler}, {No{\"e}l}, {Olsen},
  {Pieres}, {Sakowska}, {Simon}, {Stringfellow}, {Tollerud}, {Vivas}, {Walker},
  {Wechsler}, \& {Delve Collaboration}}]{Cerny2023DECam}
{Cerny}, W., {Mart{\'\i}nez-V{\'a}zquez}, C.~E., {Drlica-Wagner}, A., {et~al.}
  2023{\natexlab{a}}, \apj, 953, 1, \dodoi{10.3847/1538-4357/acdd78}

\bibitem[{{Cerny} {et~al.}(2023{\natexlab{b}}){Cerny}, {Drlica-Wagner}, {Li},
  {Pace}, {Olsen}, {No{\"e}l}, {van der Marel}, {Carlin}, {Choi}, {Erkal},
  {Geha}, {James}, {Mart{\'\i}nez-V{\'a}zquez}, {Massana}, {Medina}, {Miller},
  {Mutlu-Pakdil}, {Nidever}, {Sakowska}, {Stringfellow}, {Carballo-Bello},
  {Ferguson}, {Kuropatkin}, {Mau}, {Tollerud}, {Vivas}, \& {Delve
  Collaboration}}]{Cerny2023Delve}
{Cerny}, W., {Drlica-Wagner}, A., {Li}, T.~S., {et~al.} 2023{\natexlab{b}},
  \apjl, 953, L21, \dodoi{10.3847/2041-8213/aced84}

\bibitem[{{Chiti} {et~al.}(2021){Chiti}, {Frebel}, {Simon}, {Erkal}, {Chang},
  {Necib}, {Ji}, {Jerjen}, {Kim}, \& {Norris}}]{Chiti2021}
{Chiti}, A., {Frebel}, A., {Simon}, J.~D., {et~al.} 2021, Nature Astronomy, 5,
  392, \dodoi{10.1038/s41550-020-01285-w}

\bibitem[{{Col{\'\i}n} {et~al.}(2002){Col{\'\i}n}, {Avila-Reese}, {Valenzuela},
  \& {Firmani}}]{Colin2002}
{Col{\'\i}n}, P., {Avila-Reese}, V., {Valenzuela}, O., \& {Firmani}, C. 2002,
  \apj, 581, 777, \dodoi{10.1086/344259}

\bibitem[{{Collins} {et~al.}(2020){Collins}, {Tollerud}, {Rich}, {Ibata},
  {Martin}, {Chapman}, {Gilbert}, \& {Preston}}]{Collins2020}
{Collins}, M. L.~M., {Tollerud}, E.~J., {Rich}, R.~M., {et~al.} 2020, \mnras,
  491, 3496, \dodoi{10.1093/mnras/stz3252}

\bibitem[{{Collins} {et~al.}(2021){Collins}, {Read}, {Ibata}, {Rich}, {Martin},
  {Pe{\~n}arrubia}, {Chapman}, {Tollerud}, \& {Weisz}}]{Collins2021}
{Collins}, M. L.~M., {Read}, J.~I., {Ibata}, R.~A., {et~al.} 2021, \mnras, 505,
  5686, \dodoi{10.1093/mnras/stab1624}

\bibitem[{{Correa} {et~al.}(2022){Correa}, {Schaller}, {Ploeckinger}, {Anau
  Montel}, {Weniger}, \& {Ando}}]{Correa2022}
{Correa}, C.~A., {Schaller}, M., {Ploeckinger}, S., {et~al.} 2022, \mnras, 517,
  3045, \dodoi{10.1093/mnras/stac2830}

\bibitem[{{Drlica-Wagner} {et~al.}(2015){Drlica-Wagner}, {Bechtol}, {Rykoff},
  {Luque}, {Queiroz}, {Mao}, {Wechsler}, {Simon}, {Santiago}, {Yanny},
  {Balbinot}, {Dodelson}, {Fausti Neto}, {James}, {Li}, {Maia}, {Marshall},
  {Pieres}, {Stringer}, {Walker}, {Abbott}, {Abdalla}, {Allam},
  {Benoit-L{\'e}vy}, {Bernstein}, {Bertin}, {Brooks}, {Buckley-Geer}, {Burke},
  {Carnero Rosell}, {Carrasco Kind}, {Carretero}, {Crocce}, {da Costa},
  {Desai}, {Diehl}, {Dietrich}, {Doel}, {Eifler}, {Evrard}, {Finley},
  {Flaugher}, {Fosalba}, {Frieman}, {Gaztanaga}, {Gerdes}, {Gruen}, {Gruendl},
  {Gutierrez}, {Honscheid}, {Kuehn}, {Kuropatkin}, {Lahav}, {Martini},
  {Miquel}, {Nord}, {Ogando}, {Plazas}, {Reil}, {Roodman}, {Sako}, {Sanchez},
  {Scarpine}, {Schubnell}, {Sevilla-Noarbe}, {Smith}, {Soares-Santos},
  {Sobreira}, {Suchyta}, {Swanson}, {Tarle}, {Tucker}, {Vikram}, {Wester},
  {Zhang}, {Zuntz}, \& {DES Collaboration}}]{Drlica-Wagner2015}
{Drlica-Wagner}, A., {Bechtol}, K., {Rykoff}, E.~S., {et~al.} 2015, \apj, 813,
  109, \dodoi{10.1088/0004-637X/813/2/109}

\bibitem[{{Efstathiou}(1992)}]{Efstathiou1992}
{Efstathiou}, G. 1992, \mnras, 256, 43P, \dodoi{10.1093/mnras/256.1.43P}

\bibitem[{{Eilers} {et~al.}(2019){Eilers}, {Hogg}, {Rix}, \&
  {Ness}}]{Eilers2019}
{Eilers}, A.-C., {Hogg}, D.~W., {Rix}, H.-W., \& {Ness}, M.~K. 2019, \apj, 871,
  120, \dodoi{10.3847/1538-4357/aaf648}

\bibitem[{{Errani} \& {Navarro}(2021)}]{EN21}
{Errani}, R., \& {Navarro}, J.~F. 2021, \mnras, 505, 18,
  \dodoi{10.1093/mnras/stab1215}

\bibitem[{{Errani} {et~al.}(2022){Errani}, {Navarro}, {Ibata}, \&
  {Pe{\~n}arrubia}}]{ENIP2022}
{Errani}, R., {Navarro}, J.~F., {Ibata}, R., \& {Pe{\~n}arrubia}, J. 2022,
  \mnras, 511, 6001, \dodoi{10.1093/mnras/stac476}

\bibitem[{{Errani} {et~al.}(2023){Errani}, {Navarro}, {Pe{\~n}arrubia},
  {Famaey}, \& {Ibata}}]{ENPFI2023}
{Errani}, R., {Navarro}, J.~F., {Pe{\~n}arrubia}, J., {Famaey}, B., \& {Ibata},
  R. 2023, \mnras, 519, 384, \dodoi{10.1093/mnras/stac3499}

\bibitem[{{Errani} \& {Pe{\~n}arrubia}(2020)}]{EP20}
{Errani}, R., \& {Pe{\~n}arrubia}, J. 2020, \mnras, 491, 4591,
  \dodoi{10.1093/mnras/stz3349}

\bibitem[{{Errani} {et~al.}(2018){Errani}, {Pe{\~n}arrubia}, \&
  {Walker}}]{EPW18}
{Errani}, R., {Pe{\~n}arrubia}, J., \& {Walker}, M.~G. 2018, \mnras, 481, 5073,
  \dodoi{10.1093/mnras/sty2505}

\bibitem[{{Evans} {et~al.}(2016){Evans}, {Sanders}, \&
  {Geringer-Sameth}}]{Evans2016}
{Evans}, N.~W., {Sanders}, J.~L., \& {Geringer-Sameth}, A. 2016, \prd, 93,
  103512, \dodoi{10.1103/PhysRevD.93.103512}

\bibitem[{{Fattahi} {et~al.}(2018){Fattahi}, {Navarro}, {Frenk}, {Oman},
  {Sawala}, \& {Schaller}}]{Fattahi2018}
{Fattahi}, A., {Navarro}, J.~F., {Frenk}, C.~S., {et~al.} 2018, \mnras, 476,
  3816, \dodoi{10.1093/mnras/sty408}

\bibitem[{{Fellhauer} {et~al.}(2000){Fellhauer}, {Kroupa}, {Baumgardt}, {Bien},
  {Boily}, {Spurzem}, \& {Wassmer}}]{Fellhauer2000}
{Fellhauer}, M., {Kroupa}, P., {Baumgardt}, H., {et~al.} 2000, NA, 5, 305,
  \dodoi{10.1016/S1384-1076(00)00032-4}

\bibitem[{{Ferreira}(2021)}]{Ferreira2021}
{Ferreira}, E. G.~M. 2021, \aapr, 29, 7, \dodoi{10.1007/s00159-021-00135-6}

\bibitem[{{Ferrero} {et~al.}(2012){Ferrero}, {Abadi}, {Navarro}, {Sales}, \&
  {Gurovich}}]{Ferrero2012}
{Ferrero}, I., {Abadi}, M.~G., {Navarro}, J.~F., {Sales}, L.~V., \& {Gurovich},
  S. 2012, \mnras, 425, 2817, \dodoi{10.1111/j.1365-2966.2012.21623.x}

\bibitem[{{Frebel} \& {Norris}(2015)}]{Frebel2015}
{Frebel}, A., \& {Norris}, J.~E. 2015, \araa, 53, 631,
  \dodoi{10.1146/annurev-astro-082214-122423}

\bibitem[{{Geha} {et~al.}(2009){Geha}, {Willman}, {Simon}, {Strigari}, {Kirby},
  {Law}, \& {Strader}}]{Geha2009}
{Geha}, M., {Willman}, B., {Simon}, J.~D., {et~al.} 2009, \apj, 692, 1464,
  \dodoi{10.1088/0004-637X/692/2/1464}

\bibitem[{{Gieles} {et~al.}(2021){Gieles}, {Erkal}, {Antonini}, {Balbinot}, \&
  {Pe{\~n}arrubia}}]{Gieles2021}
{Gieles}, M., {Erkal}, D., {Antonini}, F., {Balbinot}, E., \& {Pe{\~n}arrubia},
  J. 2021, Nature Astronomy, 5, 957, \dodoi{10.1038/s41550-021-01392-2}

\bibitem[{{Gnedin}(2000)}]{Gnedin2000}
{Gnedin}, N.~Y. 2000, \apj, 542, 535, \dodoi{10.1086/317042}

\bibitem[{{Gonz{\'a}lez-Morales} {et~al.}(2017){Gonz{\'a}lez-Morales}, {Marsh},
  {Pe{\~n}arrubia}, \& {Ure{\~n}a-L{\'o}pez}}]{GonzalezMorales2017}
{Gonz{\'a}lez-Morales}, A.~X., {Marsh}, D. J.~E., {Pe{\~n}arrubia}, J., \&
  {Ure{\~n}a-L{\'o}pez}, L.~A. 2017, \mnras, 472, 1346,
  \dodoi{10.1093/mnras/stx1941}

\bibitem[{{Goodman}(2000)}]{Goodman2000}
{Goodman}, J. 2000, \na, 5, 103, \dodoi{10.1016/S1384-1076(00)00015-4}

\bibitem[{{Hansen} {et~al.}(2017){Hansen}, {Simon}, {Marshall}, {Li},
  {Carollo}, {DePoy}, {Nagasawa}, {Bernstein}, {Drlica-Wagner}, {Abdalla},
  {Allam}, {Annis}, {Bechtol}, {Benoit-L{\'e}vy}, {Brooks}, {Buckley-Geer},
  {Carnero Rosell}, {Carrasco Kind}, {Carretero}, {Cunha}, {da Costa}, {Desai},
  {Eifler}, {Fausti Neto}, {Flaugher}, {Frieman}, {Garc{\'\i}a-Bellido},
  {Gaztanaga}, {Gerdes}, {Gruen}, {Gruendl}, {Gschwend}, {Gutierrez}, {James},
  {Krause}, {Kuehn}, {Kuropatkin}, {Lahav}, {Miquel}, {Plazas}, {Romer},
  {Sanchez}, {Santiago}, {Scarpine}, {Smith}, {Soares-Santos}, {Sobreira},
  {Suchyta}, {Swanson}, {Tarle}, {Walker}, \& {DES Collaboration}}]{Hansen2017}
{Hansen}, T.~T., {Simon}, J.~D., {Marshall}, J.~L., {et~al.} 2017, \apj, 838,
  44, \dodoi{10.3847/1538-4357/aa634a}

\bibitem[{{Harris}(1996)}]{Harris1996}
{Harris}, W.~E. 1996, \aj, 112, 1487, \dodoi{10.1086/118116}

\bibitem[{{Hayashi} {et~al.}(2021){Hayashi}, {Ibe}, {Kobayashi}, {Nakayama}, \&
  {Shirai}}]{Hayashi2021}
{Hayashi}, K., {Ibe}, M., {Kobayashi}, S., {Nakayama}, Y., \& {Shirai}, S.
  2021, \prd, 103, 023017, \dodoi{10.1103/PhysRevD.103.023017}

\bibitem[{{Hernquist}(1990)}]{hernquist1990}
{Hernquist}, L. 1990, ApJ, 356, 359, \dodoi{10.1086/168845}

\bibitem[{{Hilker}(2006)}]{Hiker2006}
{Hilker}, M. 2006, \aap, 448, 171, \dodoi{10.1051/0004-6361:20054327}

\bibitem[{{Hu} {et~al.}(2000){Hu}, {Barkana}, \& {Gruzinov}}]{Hu2000}
{Hu}, W., {Barkana}, R., \& {Gruzinov}, A. 2000, Physical Review Letters, 85,
  1158, \dodoi{10.1103/PhysRevLett.85.1158}

\bibitem[{{Huang} {et~al.}(2016){Huang}, {Liu}, {Yuan}, {Xiang}, {Zhang},
  {Chen}, {Ren}, {Wang}, {Zhang}, {Hou}, {Wang}, \& {Cao}}]{Huang2016}
{Huang}, Y., {Liu}, X.~W., {Yuan}, H.~B., {et~al.} 2016, \mnras, 463, 2623,
  \dodoi{10.1093/mnras/stw2096}

\bibitem[{{Hui} {et~al.}(2017){Hui}, {Ostriker}, {Tremaine}, \&
  {Witten}}]{Hui2017}
{Hui}, L., {Ostriker}, J.~P., {Tremaine}, S., \& {Witten}, E. 2017, \prd, 95,
  043541, \dodoi{10.1103/PhysRevD.95.043541}

\bibitem[{{Inman} \& {Carney}(1987)}]{Inman1987}
{Inman}, R.~T., \& {Carney}, B.~W. 1987, \aj, 93, 1166, \dodoi{10.1086/114398}

\bibitem[{{Ji} {et~al.}(2019){Ji}, {Simon}, {Frebel}, {Venn}, \&
  {Hansen}}]{Ji2019_Gru1}
{Ji}, A.~P., {Simon}, J.~D., {Frebel}, A., {Venn}, K.~A., \& {Hansen}, T.~T.
  2019, \apj, 870, 83, \dodoi{10.3847/1538-4357/aaf3bb}

\bibitem[{{Ji} {et~al.}(2021){Ji}, {Koposov}, {Li}, {Erkal}, {Pace}, {Simon},
  {Belokurov}, {Cullinane}, {Da Costa}, {Kuehn}, {Lewis}, {Mackey}, {Shipp},
  {Simpson}, {Zucker}, {Hansen}, {Bland-Hawthorn}, \& {S5
  Collaboration}}]{Ji2021}
{Ji}, A.~P., {Koposov}, S.~E., {Li}, T.~S., {et~al.} 2021, \apj, 921, 32,
  \dodoi{10.3847/1538-4357/ac1869}

\bibitem[{{Jordi} {et~al.}(2009){Jordi}, {Grebel}, {Hilker}, {Baumgardt},
  {Frank}, {Kroupa}, {Haghi}, {C{\^o}t{\'e}}, \& {Djorgovski}}]{Jordi2009}
{Jordi}, K., {Grebel}, E.~K., {Hilker}, M., {et~al.} 2009, \aj, 137, 4586,
  \dodoi{10.1088/0004-6256/137/6/4586}

\bibitem[{{Koposov} {et~al.}(2007){Koposov}, {de Jong}, {Belokurov}, {Rix},
  {Zucker}, {Evans}, {Gilmore}, {Irwin}, \& {Bell}}]{Koposov2007}
{Koposov}, S., {de Jong}, J.~T.~A., {Belokurov}, V., {et~al.} 2007, \apj, 669,
  337, \dodoi{10.1086/521422}

\bibitem[{{Koposov} {et~al.}(2008){Koposov}, {Belokurov}, {Evans}, {Hewett},
  {Irwin}, {Gilmore}, {Zucker}, {Rix}, {Fellhauer}, {Bell}, \&
  {Glushkova}}]{Koposov2008}
{Koposov}, S., {Belokurov}, V., {Evans}, N.~W., {et~al.} 2008, \apj, 686, 279,
  \dodoi{10.1086/589911}

\bibitem[{{Kuzma} {et~al.}(2015){Kuzma}, {Da Costa}, {Keller}, \&
  {Maunder}}]{Kuzma2015}
{Kuzma}, P.~B., {Da Costa}, G.~S., {Keller}, S.~C., \& {Maunder}, E. 2015,
  \mnras, 446, 3297, \dodoi{10.1093/mnras/stu2343}

\bibitem[{{Laporte} {et~al.}(2019){Laporte}, {Agnello}, \&
  {Navarro}}]{Laporte2019}
{Laporte}, C. F.~P., {Agnello}, A., \& {Navarro}, J.~F. 2019, \mnras, 484, 245,
  \dodoi{10.1093/mnras/sty2891}

\bibitem[{{Li} {et~al.}(2021){Li}, {Hammer}, {Babusiaux}, {Pawlowski}, {Yang},
  {Arenou}, {Du}, \& {Wang}}]{LiHammer2021}
{Li}, H., {Hammer}, F., {Babusiaux}, C., {et~al.} 2021, \apj, 916, 8,
  \dodoi{10.3847/1538-4357/ac0436}

\bibitem[{{Lovell} {et~al.}(2014){Lovell}, {Frenk}, {Eke}, {Jenkins}, {Gao}, \&
  {Theuns}}]{Lovell2014}
{Lovell}, M.~R., {Frenk}, C.~S., {Eke}, V.~R., {et~al.} 2014, \mnras, 439, 300,
  \dodoi{10.1093/mnras/stt2431}

\bibitem[{{Ludlow} {et~al.}(2016){Ludlow}, {Bose}, {Angulo}, {Wang},
  {Hellwing}, {Navarro}, {Cole}, \& {Frenk}}]{Ludlow2016}
{Ludlow}, A.~D., {Bose}, S., {Angulo}, R.~E., {et~al.} 2016, \mnras, 460, 1214,
  \dodoi{10.1093/mnras/stw1046}

\bibitem[{{Marsh} \& {Pop}(2015)}]{Marsh2015}
{Marsh}, D. J.~E., \& {Pop}, A.-R. 2015, \mnras, 451, 2479,
  \dodoi{10.1093/mnras/stv1050}

\bibitem[{{Marshall} {et~al.}(2019){Marshall}, {Hansen}, {Simon}, {Li},
  {Bernstein}, {Kuehn}, {Pace}, {DePoy}, {Palmese}, {Pieres}, {Strigari},
  {Drlica-Wagner}, {Bechtol}, {Lidman}, {Nagasawa}, {Bertin}, {Brooks},
  {Buckley-Geer}, {Burke}, {Carnero Rosell}, {Carrasco Kind}, {Carretero},
  {Cunha}, {D'Andrea}, {da Costa}, {De Vicente}, {Desai}, {Doel}, {Eifler},
  {Flaugher}, {Fosalba}, {Frieman}, {Garc{\'\i}a-Bellido}, {Gaztanaga},
  {Gerdes}, {Gruendl}, {Gschwend}, {Gutierrez}, {Hartley}, {Hollowood},
  {Honscheid}, {Hoyle}, {James}, {Kuropatkin}, {Maia}, {Menanteau}, {Miller},
  {Miquel}, {Plazas}, {Sanchez}, {Santiago}, {Scarpine}, {Schubnell},
  {Serrano}, {Sevilla-Noarbe}, {Smith}, {Soares-Santos}, {Suchyta}, {Swanson},
  {Tarle}, {Wester}, \& {DES Collaboration}}]{Marshall2019}
{Marshall}, J.~L., {Hansen}, T., {Simon}, J.~D., {et~al.} 2019, \apj, 882, 177,
  \dodoi{10.3847/1538-4357/ab3653}

\bibitem[{{Martin} {et~al.}(2013){Martin}, {Ibata}, {McConnachie}, {Mackey},
  {Ferguson}, {Irwin}, {Lewis}, \& {Fardal}}]{Martin2013}
{Martin}, N.~F., {Ibata}, R.~A., {McConnachie}, A.~W., {et~al.} 2013, \apj,
  776, 80, \dodoi{10.1088/0004-637X/776/2/80}

\bibitem[{{Martin} {et~al.}(2016){Martin}, {Geha}, {Ibata}, {Collins},
  {Laevens}, {Bell}, {Rix}, {Ferguson}, {Chambers}, {Wainscoat}, \&
  {Waters}}]{Martin2016_Dra2}
{Martin}, N.~F., {Geha}, M., {Ibata}, R.~A., {et~al.} 2016, \mnras, 458, L59,
  \dodoi{10.1093/mnrasl/slw013}

\bibitem[{{Mau} {et~al.}(2020){Mau}, {Cerny}, {Pace}, {Choi}, {Drlica-Wagner},
  {Santana-Silva}, {Riley}, {Erkal}, {Stringfellow}, {Adam{\'o}w}, {Carlin},
  {Gruendl}, {Hernandez-Lang}, {Kuropatkin}, {Li}, {Mart{\'\i}nez-V{\'a}zquez},
  {Morganson}, {Mutlu-Pakdil}, {Neilsen}, {Nidever}, {Olsen}, {Sand},
  {Tollerud}, {Tucker}, {Yanny}, {Zenteno}, {Allam}, {Barkhouse}, {Bechtol},
  {Bell}, {Balaji}, {Crnojevi{\'c}}, {Esteves}, {Ferguson}, {Gallart},
  {Hughes}, {James}, {Jethwa}, {Johnson}, {Kuehn}, {Majewski}, {Mao},
  {Massana}, {McNanna}, {Monachesi}, {Nadler}, {No{\"e}l}, {Palmese},
  {Paz-Chinchon}, {Pieres}, {Sanchez}, {Shipp}, {Simon}, {Soares-Santos},
  {Tavangar}, {van der Marel}, {Vivas}, {Walker}, \& {Wechsler}}]{Mau2020}
{Mau}, S., {Cerny}, W., {Pace}, A.~B., {et~al.} 2020, \apj, 890, 136,
  \dodoi{10.3847/1538-4357/ab6c67}

\bibitem[{{McConnachie}(2012)}]{McConnachie2012}
{McConnachie}, A.~W. 2012, \aj, 144, 4, \dodoi{10.1088/0004-6256/144/1/4}

\bibitem[{{McConnachie} \& {C{\^o}t{\'e}}(2010)}]{McConnachieCote2010}
{McConnachie}, A.~W., \& {C{\^o}t{\'e}}, P. 2010, \apjl, 722, L209,
  \dodoi{10.1088/2041-8205/722/2/L209}

\bibitem[{{McConnachie} \& {Venn}(2020{\natexlab{a}})}]{McConnachieVenn2020b}
{McConnachie}, A.~W., \& {Venn}, K.~A. 2020{\natexlab{a}}, Research Notes of
  the American Astronomical Society, 4, 229, \dodoi{10.3847/2515-5172/abd18b}

\bibitem[{{McConnachie} \& {Venn}(2020{\natexlab{b}})}]{McConnachieVenn2020a}
---. 2020{\natexlab{b}}, \aj, 160, 124, \dodoi{10.3847/1538-3881/aba4ab}

\bibitem[{{McMillan}(2011)}]{McMillan2011}
{McMillan}, P.~J. 2011, \mnras, 414, 2446,
  \dodoi{10.1111/j.1365-2966.2011.18564.x}

\bibitem[{{Minor} {et~al.}(2010){Minor}, {Martinez}, {Bullock}, {Kaplinghat},
  \& {Trainor}}]{MinorQuinn2010}
{Minor}, Q.~E., {Martinez}, G., {Bullock}, J., {Kaplinghat}, M., \& {Trainor},
  R. 2010, \apj, 721, 1142, \dodoi{10.1088/0004-637X/721/2/1142}

\bibitem[{{Miyamoto} \& {Nagai}(1975)}]{miyamoto1975}
{Miyamoto}, M., \& {Nagai}, R. 1975, PASJ, 27, 533

\bibitem[{{Mu{\~n}oz} {et~al.}(2012){Mu{\~n}oz}, {Geha}, {C{\^o}t{\'e}},
  {Vargas}, {Santana}, {Stetson}, {Simon}, \& {Djorgovski}}]{Munoz2012}
{Mu{\~n}oz}, R.~R., {Geha}, M., {C{\^o}t{\'e}}, P., {et~al.} 2012, \apjl, 753,
  L15, \dodoi{10.1088/2041-8205/753/1/L15}

\bibitem[{{Nadler} {et~al.}(2021){Nadler}, {Drlica-Wagner}, {Bechtol}, {Mau},
  {Wechsler}, {Gluscevic}, {Boddy}, {Pace}, {Li}, {McNanna}, {Riley},
  {Garc{\'\i}a-Bellido}, {Mao}, {Green}, {Burke}, {Peter}, {Jain}, {Abbott},
  {Aguena}, {Allam}, {Annis}, {Avila}, {Brooks}, {Carrasco Kind}, {Carretero},
  {Costanzi}, {da Costa}, {De Vicente}, {Desai}, {Diehl}, {Doel}, {Everett},
  {Evrard}, {Flaugher}, {Frieman}, {Gerdes}, {Gruen}, {Gruendl}, {Gschwend},
  {Gutierrez}, {Hinton}, {Honscheid}, {Huterer}, {James}, {Krause}, {Kuehn},
  {Kuropatkin}, {Lahav}, {Maia}, {Marshall}, {Menanteau}, {Miquel}, {Palmese},
  {Paz-Chinch{\'o}n}, {Plazas}, {Romer}, {Sanchez}, {Scarpine}, {Serrano},
  {Sevilla-Noarbe}, {Smith}, {Soares-Santos}, {Suchyta}, {Swanson}, {Tarle},
  {Tucker}, {Walker}, {Wester}, \& {DES Collaboration}}]{Nadler2021}
{Nadler}, E.~O., {Drlica-Wagner}, A., {Bechtol}, K., {et~al.} 2021, \prl, 126,
  091101, \dodoi{10.1103/PhysRevLett.126.091101}

\bibitem[{{Navarro} {et~al.}(1996){Navarro}, {Frenk}, \&
  {White}}]{Navarro1996a}
{Navarro}, J.~F., {Frenk}, C.~S., \& {White}, S. D.~M. 1996, \apj, 462, 563,
  \dodoi{10.1086/177173}

\bibitem[{{Navarro} {et~al.}(1997){Navarro}, {Frenk}, \& {White}}]{Navarro1997}
{Navarro}, J.~F., {Frenk}, C.~S., \& {White}, S.~D.~M. 1997, ApJ, 490, 493,
  \dodoi{10.1086/304888}

\bibitem[{{Nishikawa} {et~al.}(2020){Nishikawa}, {Boddy}, \&
  {Kaplinghat}}]{Nishikawa2020}
{Nishikawa}, H., {Boddy}, K.~K., \& {Kaplinghat}, M. 2020, \prd, 101, 063009,
  \dodoi{10.1103/PhysRevD.101.063009}

\bibitem[{{Pace} \& {Li}(2019)}]{Pace2019}
{Pace}, A.~B., \& {Li}, T.~S. 2019, \apj, 875, 77,
  \dodoi{10.3847/1538-4357/ab0aee}

\bibitem[{{Pace} \& {Strigari}(2019)}]{PaceStrigari2019}
{Pace}, A.~B., \& {Strigari}, L.~E. 2019, \mnras, 482, 3480,
  \dodoi{10.1093/mnras/sty2839}

\bibitem[{{Pe{\~n}arrubia} {et~al.}(2010){Pe{\~n}arrubia}, {Benson}, {Walker},
  {Gilmore}, {McConnachie}, \& {Mayer}}]{Penarrubia2010}
{Pe{\~n}arrubia}, J., {Benson}, A.~J., {Walker}, M.~G., {et~al.} 2010, MNRAS,
  406, 1290, \dodoi{10.1111/j.1365-2966.2010.16762.x}

\bibitem[{{Pe{\~n}arrubia} {et~al.}(2008){Pe{\~n}arrubia}, {Navarro}, \&
  {McConnachie}}]{Penarrubia2008}
{Pe{\~n}arrubia}, J., {Navarro}, J.~F., \& {McConnachie}, A.~W. 2008, ApJ, 673,
  226, \dodoi{10.1086/523686}

\bibitem[{{Pe{\~n}arrubia} {et~al.}(2012){Pe{\~n}arrubia}, {Pontzen}, {Walker},
  \& {Koposov}}]{Penarrubia2012}
{Pe{\~n}arrubia}, J., {Pontzen}, A., {Walker}, M.~G., \& {Koposov}, S.~E. 2012,
  \apjl, 759, L42, \dodoi{10.1088/2041-8205/759/2/L42}

\bibitem[{{Pereira-Wilson} {et~al.}(2023){Pereira-Wilson}, {Navarro},
  {Ben{\'\i}tez-Llambay}, \& {Santos-Santos}}]{Pereira-Wilson23}
{Pereira-Wilson}, M., {Navarro}, J.~F., {Ben{\'\i}tez-Llambay}, A., \&
  {Santos-Santos}, I. 2023, \mnras, 519, 1425, \dodoi{10.1093/mnras/stac3633}

\bibitem[{{Planck Collaboration} {et~al.}(2020){Planck Collaboration},
  {Aghanim}, {Akrami}, {Ashdown}, {Aumont}, {Baccigalupi}, {Ballardini},
  {Banday}, {Barreiro}, {Bartolo}, {Basak}, {Battye}, {Benabed}, {Bernard},
  {Bersanelli}, {Bielewicz}, {Bock}, {Bond}, {Borrill}, {Bouchet}, {Boulanger},
  {Bucher}, {Burigana}, {Butler}, {Calabrese}, {Cardoso}, {Carron},
  {Challinor}, {Chiang}, {Chluba}, {Colombo}, {Combet}, {Contreras}, {Crill},
  {Cuttaia}, {de Bernardis}, {de Zotti}, {Delabrouille}, {Delouis}, {Di
  Valentino}, {Diego}, {Dor{\'e}}, {Douspis}, {Ducout}, {Dupac}, {Dusini},
  {Efstathiou}, {Elsner}, {En{\ss}lin}, {Eriksen}, {Fantaye}, {Farhang},
  {Fergusson}, {Fernandez-Cobos}, {Finelli}, {Forastieri}, {Frailis},
  {Fraisse}, {Franceschi}, {Frolov}, {Galeotta}, {Galli}, {Ganga},
  {G{\'e}nova-Santos}, {Gerbino}, {Ghosh}, {Gonz{\'a}lez-Nuevo}, {G{\'o}rski},
  {Gratton}, {Gruppuso}, {Gudmundsson}, {Hamann}, {Handley}, {Hansen},
  {Herranz}, {Hildebrandt}, {Hivon}, {Huang}, {Jaffe}, {Jones}, {Karakci},
  {Keih{\"a}nen}, {Keskitalo}, {Kiiveri}, {Kim}, {Kisner}, {Knox},
  {Krachmalnicoff}, {Kunz}, {Kurki-Suonio}, {Lagache}, {Lamarre}, {Lasenby},
  {Lattanzi}, {Lawrence}, {Le Jeune}, {Lemos}, {Lesgourgues}, {Levrier},
  {Lewis}, {Liguori}, {Lilje}, {Lilley}, {Lindholm}, {L{\'o}pez-Caniego},
  {Lubin}, {Ma}, {Mac{\'\i}as-P{\'e}rez}, {Maggio}, {Maino}, {Mandolesi},
  {Mangilli}, {Marcos-Caballero}, {Maris}, {Martin}, {Martinelli},
  {Mart{\'\i}nez-Gonz{\'a}lez}, {Matarrese}, {Mauri}, {McEwen}, {Meinhold},
  {Melchiorri}, {Mennella}, {Migliaccio}, {Millea}, {Mitra},
  {Miville-Desch{\^e}nes}, {Molinari}, {Montier}, {Morgante}, {Moss}, {Natoli},
  {N{\o}rgaard-Nielsen}, {Pagano}, {Paoletti}, {Partridge}, {Patanchon},
  {Peiris}, {Perrotta}, {Pettorino}, {Piacentini}, {Polastri}, {Polenta},
  {Puget}, {Rachen}, {Reinecke}, {Remazeilles}, {Renzi}, {Rocha}, {Rosset},
  {Roudier}, {Rubi{\~n}o-Mart{\'\i}n}, {Ruiz-Granados}, {Salvati}, {Sandri},
  {Savelainen}, {Scott}, {Shellard}, {Sirignano}, {Sirri}, {Spencer},
  {Sunyaev}, {Suur-Uski}, {Tauber}, {Tavagnacco}, {Tenti}, {Toffolatti},
  {Tomasi}, {Trombetti}, {Valenziano}, {Valiviita}, {Van Tent}, {Vibert},
  {Vielva}, {Villa}, {Vittorio}, {Wandelt}, {Wehus}, {White}, {White},
  {Zacchei}, \& {Zonca}}]{Planck2020}
{Planck Collaboration}, {Aghanim}, N., {Akrami}, Y., {et~al.} 2020, \aap, 641,
  A6, \dodoi{10.1051/0004-6361/201833910}

\bibitem[{{Quinn} {et~al.}(1996){Quinn}, {Katz}, \& {Efstathiou}}]{Quinn1996}
{Quinn}, T., {Katz}, N., \& {Efstathiou}, G. 1996, \mnras, 278, L49,
  \dodoi{10.1093/mnras/278.4.L49}

\bibitem[{{Safarzadeh} \& {Spergel}(2020)}]{Safarzadeh2020}
{Safarzadeh}, M., \& {Spergel}, D.~N. 2020, \apj, 893, 21,
  \dodoi{10.3847/1538-4357/ab7db2}

\bibitem[{{Sales} {et~al.}(2022){Sales}, {Wetzel}, \& {Fattahi}}]{Sales2022}
{Sales}, L.~V., {Wetzel}, A., \& {Fattahi}, A. 2022, Nature Astronomy, 6, 897,
  \dodoi{10.1038/s41550-022-01689-w}

\bibitem[{{Schive} {et~al.}(2014{\natexlab{a}}){Schive}, {Chiueh}, \&
  {Broadhurst}}]{Schive2014}
{Schive}, H.-Y., {Chiueh}, T., \& {Broadhurst}, T. 2014{\natexlab{a}}, Nature
  Physics, 10, 496, \dodoi{10.1038/nphys2996}

\bibitem[{{Schive} {et~al.}(2014{\natexlab{b}}){Schive}, {Liao}, {Woo}, {Wong},
  {Chiueh}, {Broadhurst}, \& {Hwang}}]{Schive2014b}
{Schive}, H.-Y., {Liao}, M.-H., {Woo}, T.-P., {et~al.} 2014{\natexlab{b}},
  \prl, 113, 261302, \dodoi{10.1103/PhysRevLett.113.261302}

\bibitem[{{Sch{\"o}nrich} {et~al.}(2010){Sch{\"o}nrich}, {Binney}, \&
  {Dehnen}}]{SB2010}
{Sch{\"o}nrich}, R., {Binney}, J., \& {Dehnen}, W. 2010, \mnras, 403, 1829,
  \dodoi{10.1111/j.1365-2966.2010.16253.x}

\bibitem[{{Silverman} {et~al.}(2023){Silverman}, {Bullock}, {Kaplinghat},
  {Robles}, \& {Valli}}]{Silverman2023}
{Silverman}, M., {Bullock}, J.~S., {Kaplinghat}, M., {Robles}, V.~H., \&
  {Valli}, M. 2023, \mnras, 518, 2418, \dodoi{10.1093/mnras/stac3232}

\bibitem[{{Simon}(2018)}]{Simon2018}
{Simon}, J.~D. 2018, \apj, 863, 89, \dodoi{10.3847/1538-4357/aacdfb}

\bibitem[{{Simon}(2019)}]{Simon2019Review}
---. 2019, \araa, 57, 375, \dodoi{10.1146/annurev-astro-091918-104453}

\bibitem[{{Simon} \& {Geha}(2007)}]{Simon2007}
{Simon}, J.~D., \& {Geha}, M. 2007, \apj, 670, 313, \dodoi{10.1086/521816}

\bibitem[{{Smith} {et~al.}(2024){Smith}, {Cerny}, {Hayes}, {Sestito}, {Jensen},
  {McConnachie}, {Geha}, {Navarro}, {Li}, {Cuillandre}, {Errani}, {Chambers},
  {Gwyn}, {Hammer}, {Hudson}, {Magnier}, \& {Martin}}]{Smith2024}
{Smith}, S. E.~T., {Cerny}, W., {Hayes}, C.~R., {et~al.} 2024, \apj, 961, 92,
  \dodoi{10.3847/1538-4357/ad0d9f}

\bibitem[{{Spitzer}(1987)}]{Spitzer1987book}
{Spitzer}, L. 1987, {Dynamical evolution of globular clusters}

\bibitem[{{Spurzem} \& {Kamlah}(2023)}]{Spurzem2023}
{Spurzem}, R., \& {Kamlah}, A. 2023, Living Reviews in Computational
  Astrophysics, 9, 3, \dodoi{10.1007/s41115-023-00018-w}

\bibitem[{{Taibi} {et~al.}(2020){Taibi}, {Battaglia}, {Rejkuba}, {Leaman},
  {Kacharov}, {Iorio}, {Jablonka}, \& {Zoccali}}]{Taibi2020}
{Taibi}, S., {Battaglia}, G., {Rejkuba}, M., {et~al.} 2020, \aap, 635, A152,
  \dodoi{10.1051/0004-6361/201937240}

\bibitem[{{Torrealba} {et~al.}(2019){Torrealba}, {Belokurov}, \&
  {Koposov}}]{Torrealba2019clusters}
{Torrealba}, G., {Belokurov}, V., \& {Koposov}, S.~E. 2019, \mnras, 484, 2181,
  \dodoi{10.1093/mnras/stz071}

\bibitem[{{Tulin} \& {Yu}(2018)}]{Tulin2018}
{Tulin}, S., \& {Yu}, H.-B. 2018, \physrep, 730, 1,
  \dodoi{10.1016/j.physrep.2017.11.004}

\bibitem[{{Turner} {et~al.}(2021){Turner}, {Lovell}, {Zavala}, \&
  {Vogelsberger}}]{Turner2021}
{Turner}, H.~C., {Lovell}, M.~R., {Zavala}, J., \& {Vogelsberger}, M. 2021,
  \mnras, 505, 5327, \dodoi{10.1093/mnras/stab1725}

\bibitem[{{Vogelsberger} {et~al.}(2014){Vogelsberger}, {Zavala}, {Simpson}, \&
  {Jenkins}}]{Vogelsberger2014}
{Vogelsberger}, M., {Zavala}, J., {Simpson}, C., \& {Jenkins}, A. 2014, \mnras,
  444, 3684, \dodoi{10.1093/mnras/stu1713}

\bibitem[{{Walker} {et~al.}(2011){Walker}, {Combet}, {Hinton}, {Maurin}, \&
  {Wilkinson}}]{Walker2011_Jfact}
{Walker}, M.~G., {Combet}, C., {Hinton}, J.~A., {Maurin}, D., \& {Wilkinson},
  M.~I. 2011, \apjl, 733, L46, \dodoi{10.1088/2041-8205/733/2/L46}

\bibitem[{{Walsh} {et~al.}(2009){Walsh}, {Willman}, \& {Jerjen}}]{Walsh2009}
{Walsh}, S.~M., {Willman}, B., \& {Jerjen}, H. 2009, \aj, 137, 450,
  \dodoi{10.1088/0004-6256/137/1/450}

\bibitem[{{White} \& {Rees}(1978)}]{WhiteRees1978}
{White}, S.~D.~M., \& {Rees}, M.~J. 1978, \mnras, 183, 341,
  \dodoi{10.1093/mnras/183.3.341}

\bibitem[{{Willman} \& {Strader}(2012)}]{WillmanStrader2012}
{Willman}, B., \& {Strader}, J. 2012, \aj, 144, 76,
  \dodoi{10.1088/0004-6256/144/3/76}

\bibitem[{{Wolf} {et~al.}(2010){Wolf}, {Martinez}, {Bullock}, {Kaplinghat},
  {Geha}, {Mu{\~n}oz}, {Simon}, \& {Avedo}}]{Wolf2010}
{Wolf}, J., {Martinez}, G.~D., {Bullock}, J.~S., {et~al.} 2010, MNRAS, 406,
  1220, \dodoi{10.1111/j.1365-2966.2010.16753.x}

\bibitem[{{Woo} {et~al.}(2008){Woo}, {Courteau}, \& {Dekel}}]{Woo2008}
{Woo}, J., {Courteau}, S., \& {Dekel}, A. 2008, \mnras, 390, 1453,
  \dodoi{10.1111/j.1365-2966.2008.13770.x}

\bibitem[{{Zavala} {et~al.}(2013){Zavala}, {Vogelsberger}, \&
  {Walker}}]{Zavala2013}
{Zavala}, J., {Vogelsberger}, M., \& {Walker}, M.~G. 2013, \mnras, 431, L20,
  \dodoi{10.1093/mnrasl/sls053}

\bibitem[{{Zeng} {et~al.}(2022){Zeng}, {Peter}, {Du}, {Benson}, {Kim}, {Jiang},
  {Cyr-Racine}, \& {Vogelsberger}}]{Zeng2022}
{Zeng}, Z.~C., {Peter}, A. H.~G., {Du}, X., {et~al.} 2022, \mnras, 513, 4845,
  \dodoi{10.1093/mnras/stac1094}

\bibitem[{{Zeng} {et~al.}(2023){Zeng}, {Peter}, {Du}, {Yang}, {Benson},
  {Cyr-Racine}, {Jiang}, {Mace}, \& {Metcalf}}]{Zeng2023}
---. 2023, arXiv e-prints, arXiv:2310.09910, \dodoi{10.48550/arXiv.2310.09910}

\end{thebibliography}

\end{document}